\documentclass[12pt]{article}

\usepackage{amsfonts}
\usepackage{amsmath}
\usepackage{amsthm}
\usepackage[driverfallback=dvipdfm]{hyperref}
\usepackage{cite}
\usepackage{epsfig}
\usepackage{latexsym}
\usepackage{paralist}
\usepackage{fancyhdr}
\usepackage{graphicx}
\numberwithin{equation}{section}
\usepackage[vcentermath]{youngtab}
\usepackage{young}
\usepackage{ytableau}
\usepackage{etex}
\usepackage{braket}
\usepackage{float}
\usepackage{pict2e}   
\usepackage{animate}  
\usepackage{multirow}
\usepackage{bigdelim}
\usepackage{fancybox}
\usepackage{cals}

\def\be{\begin{equation}}
\def\ee{\end{equation}}
\def\bea{\begin{eqnarray}}
\def\eea{\end{eqnarray}}

\renewcommand{\thefootnote}{\fnsymbol{footnote}}

\begin{document}

\hfuzz=100pt
\title{{\Large \bf{Confinement in 3d $\mathcal{N}=2$ $Spin(N)$ gauge theories\\ with vector and spinor matters} }}
\date{}
\author{ Keita Nii$^a$\footnote{nii@itp.unibe.ch}
%, 
%Yuta Sekiguchi$^a$\footnote{yuta@itp.unibe.ch}
% and others$^{c,d}$
}
\date{\today}

\maketitle

\thispagestyle{fancy}
%\rhead{****-**-**}
\cfoot{}
\renewcommand{\headrulewidth}{0.0pt}

\vspace*{-1cm}
\begin{center}
%  \spa{0.5} \\
$^{a}${{\it Albert Einstein Center for Fundamental Physics }}
\\{{\it Institute for Theoretical Physics
}}
\\ {{\it University of Bern}}  
\\{{\it  Sidlerstrasse 5, CH-3012 Bern, Switzerland}}
%\\ {{\it  }}
 % \spa{0.5} \\
%$^b${{\it Department of Physics}}
%\\ {{\it Nagoya University, Nagoya 464-8602, Japan}}
%\spa{0.5}  \\
%$^c${{\it  }}
%\\ {{\it }}
%\spa{0.5}  \\
%$^d${{\it }}

\end{center}

\begin{abstract}
We present various confinement phases of three-dimensional $\mathcal{N}=2$ $Spin(N)$ gauge theories with vector and spinor matters. The quantum Coulomb branch of the moduli space of vacua is drastically changed when the rank of the gauge group and the matter contents are changed. In many examples, the Coulomb branch is one- or two-dimensional but its interpretation varies. In some examples, the Coulomb branch becomes three-dimensional and we need to introduce a ``dressed'' Coulomb branch operator. 
\end{abstract}

\renewcommand{\thefootnote}{\arabic{footnote}}
\setcounter{footnote}{0}

\newpage
\tableofcontents 
%\clearpage

%%%%%%%%%%%%%%%%%%%%%%%%%%%%%%%%%%%%%%%%%%%%%%%%%%%%%%%%%%
%%%%%%%%%%%%%%%%%%%%%%%%%%%%%%%%%%%%%%%%%%%%%%%%%%%%%%%%%%
%%%%%%%%%%%%%%%%%%%%%%%%%%%%%%%%%%%%%%%%%%%%%%%%%%%%%%%%%%
\section{Introduction}
%%%%%%%%%%%%%%%%%%%%%%%%%%%%%%%%%%%%%%%%%%%%%%%%%%%%%%%%%%
%%%%%%%%%%%%%%%%%%%%%%%%%%%%%%%%%%%%%%%%%%%%%%%%%%%%%%%%%%
%%%%%%%%%%%%%%%%%%%%%%%%%%%%%%%%%%%%%%%%%%%%%%%%%%%%%%%%%%

%[Introduction to confinement]
Strongly-coupled gauge theories exhibit various phases depending on the gauge group, matter contents, spacetime dimensions, and so on. When we increase the number of dynamical matters, the theory flows to an IR-free phase. On the other hand, when reducing the dynamical matters, the theory becomes strongly-coupled and non-perturbative.
Among various strongly-coupled phases, the confinement phase is a most fascinating one since our world is described by QCD which is actually confining. The low-energy dynamics of the confining gauge theories  
is described by mesons and baryons and exhibits chiral symmetry breaking. We cannot see dynamical quarks as low-energy asymptotic states.

%[SUSY confinement/SUSY advantage]
In supersymmetric gauge theories, there is a very special class of the confinement phases, which is known as ``s-confinement''. Usually, confinement appears, being accompanied by some symmetry breaking, such as chiral symmetry breaking. However, the SUSY gauge theories sometimes show confinement without any symmetry breaking at the origin of the moduli space of vacua. This is called ``s-confinement'' \cite{Csaki:1996sm}. In addition to this special property, the supersymmetry allows us to exactly study the non-perturbative dynamics of the gauge theory because of non-renormalization theorems and the holomorphy \cite{Seiberg:1994bz, Seiberg:1994pq}. 
In 4d, the s-confinement phases are classified in \cite{Csaki:1996sm, Csaki:1996zb} for classical and exceptional gauge groups while the corresponding 3d analysis is not completely performed.

%[In this paper]
In this paper, we study the s-confinement phases of the 3d $\mathcal{N}=2$ supersymmetric $Spin(N)$ gauge theory with vector matters and spinor matters. The 3d SUSY gauge theories contain Higgs and Coulomb branches of the moduli spaces of vacua. In general, the Coulomb branch is drastically modified and different from the classical picture. 
%3dspin(7)
In \cite{Nii:2018tnd}, we studied the 3d $\mathcal{N}=2$ $Spin(7)$ gauge theory with vector and spinor matters. We found that the Coulomb moduli space is one- or two-dimensional depending on the matter contents and also found various s-confinement phases. These phases were beautifully connected to the quantum-deformed moduli space of the 4d $\mathcal{N}=1$ $Spin(7)$ gauge theory via a non-perturbative superpotential which is generated by the twisted monopoles. In this paper, we will find the similar confinement phases for the $Spin(N) ~ (N>7)$ cases and discuss that the Coulomb moduli space is more complicated and in some cases we need three coordinates for describing it. We will give a systematic way of studying the quantum Coulomb branch and the 3d s-confinement phases. These confinement phases are also connected with the 4d quantum-deformed moduli spaces \cite{Grinstein:1997zv, Grinstein:1998bu}.

%[Organization]
The rest of this paper is organized as follows. 
In Section 2, we briefly review the Coulomb branch operators which were studied in \cite{Aharony:2011ci, Aharony:2013kma}. 
In Section 3, 4, 5, 6, 7, 8 and 9, we study the 3d $\mathcal{N}=2$ $Spin(N)~( 8 \le N \le 14)$ gauge theory with vector and spinor matters. We will give a detailed analysis of the quantum Coulomb branch for each rank. 
In Section 10, we will summarize our results and comment on future directions.

%%%%%%%%%%%%%%%%%%%%%%%%%%%%%%%%%%%%%%%%%%%%%%%%%%%%%%%%%%
%%%%%%%%%%%%%%%%%%%%%%%%%%%%%%%%%%%%%%%%%%%%%%%%%%%%%%%%%%
%%%%%%%%%%%%%%%%%%%%%%%%%%%%%%%%%%%%%%%%%%%%%%%%%%%%%%%%%%
\section{Coulomb branch in $Spin(N)$ theories}
%%%%%%%%%%%%%%%%%%%%%%%%%%%%%%%%%%%%%%%%%%%%%%%%%%%%%%%%%%
%%%%%%%%%%%%%%%%%%%%%%%%%%%%%%%%%%%%%%%%%%%%%%%%%%%%%%%%%%
%%%%%%%%%%%%%%%%%%%%%%%%%%%%%%%%%%%%%%%%%%%%%%%%%%%%%%%%%%
In this section, we will briefly review some Coulomb branch operators of the 3d $\mathcal{N}=2$ $Spin(N)$ gauge theory. These were studied in \cite{Aharony:2011ci, Aharony:2013kma} for the cases where the theory contains only the vector matters. In \cite{Nii:2018tnd}, we studied these operators in the $Spin(7)$ theory with vector and spinor matters. In these examples, we learned that almost all the classical Coulomb branches are lifted and the quantum Coulomb moduli space is described by only a few operators. Here, we review these operators and explain why these directions can remain massless.

%[Y coordinate]
For theories with only vector matters, the classical Coulomb branch where the gauge group is broken as
\begin{align}
so(N) \rightarrow so(N-2) \times u(1),
\end{align}
may remain massless and the other directions are all lifted \cite{Aharony:2011ci}. We denote this operator as $Y$. For the explicit form of this operator in terms of the fundamental monopoles, see  \cite{Aharony:2011ci, Aharony:2013kma}.
\if0
%
\begin{align}
Y_1 &\simeq \exp{\sigma_1 -\sigma_2 } \\
&\vdots \\
Y_{n-1}  &\simeq \exp( \sigma_{n-1} -\sigma_n) \\
Y_{n} &\simeq  \begin{cases}
\exp(\sigma_{n-1} +\sigma_n)  & \mbox{for}~~N=2n \\
\exp(2\sigma_n)   &   \mbox{for}~~N=2n+1
 \end{cases}
\end{align}
%
this operator is defined
%
\begin{align}
Y =  \begin{cases}
    Y_1^2 Y_2^2 \cdots Y_{n-2}^2 Y_{n-1} Y_n & \mbox{for}~~N=2n \\
    Y_1^2 Y_2^2 \cdots Y_{n-1}^2 Y_n   &   \mbox{for}~~N=2n+1
  \end{cases}
\end{align}
%
\fi
Along this branch, the spinor matters are all massive while the vector matters reduce to the massless vector representations of the unbroken $Spin(N-2)$. When the number of the vector representations of $Spin(N-2)$ is less than $N-4$, there is no supersymmetric stable vacuum. Hence, the theories with $N_v \ge N-4$ vector matters can have this Coulomb branch. For $N_v < N-4$, this direction cannot be flat. The $Spin(N)$ theory only with spinor matters cannot have this branch since the low-energy $Spin(N-2)$ theory has no dynamical matter and its vacuum is unstable due to the monopole superpotential.

The second Coulomb branch $Z$ appears when the $Spin(N)$ theory includes spinor matters or when we put the 4d $\mathcal{N}=1$ $Spin(N)$ theory on a circle \cite{Davies:1999uw, Davies:2000nw,Aharony:2011ci, Aharony:2013kma, Nii:2018tnd}. This operator corresponds to the gauge symmetry breaking
\begin{align}
so(N) \rightarrow so(N-4) \times su(2) \times u(1).
\end{align}
Along this breaking, the remaining massless components of the spinor representations are charged under the $Spin(N-4) \times SU(2)$ and chargeless under the $U(1)$. Therefore, the low-energy $Spin(N-4) \times SU(2)$ theory may have a stable SUSY vacuum because of the massless dynamical fields. If we consider this branch for the theory only with vector matters, the low-energy $SU(2)$ theory has no massless charged field and the supersymmetry is broken  by the monopole superpotential of the $SU(2)$. 
As a result, this branch $Z$ is available only for the theories with spinor matters. When we consider the 4d theory on a circle, there are two non-perturbative effects from monopoles and twisted monopoles and this flat direction is quantum-mechanically stable even for the pure SYM without a matter.

In the following sections, we will study the 3d $\mathcal{N}=2$ $Spin(N)$ gauge theories with $7 < N< 15$, where we will find that the quantum Coulomb branch becomes more richer and we need additional operators to parametrize the additional Coulomb branches. Since the corresponding breaking patterns depend on the rank of the gauge group, we will give a case-by-case analysis in what follows. See \cite{Slansky:1981yr, Georgi:1982jb, Feger:2012bs, Yamatsu:2015npn} for various branching rules.

%%%%%%%%%%%%%%%%%%%%%%%%%%%%%%%%%%%%%%%%%%%%%%%%%%%%%%%%%%
%%%%%%%%%%%%%%%%%%%%%%%%%%%%%%%%%%%%%%%%%%%%%%%%%%%%%%%%%%
%%%%%%%%%%%%%%%%%%%%%%%%%%%%%%%%%%%%%%%%%%%%%%%%%%%%%%%%%%
\section{$Spin(8)$ theories}
%%%%%%%%%%%%%%%%%%%%%%%%%%%%%%%%%%%%%%%%%%%%%%%%%%%%%%%%%%
%%%%%%%%%%%%%%%%%%%%%%%%%%%%%%%%%%%%%%%%%%%%%%%%%%%%%%%%%%
%%%%%%%%%%%%%%%%%%%%%%%%%%%%%%%%%%%%%%%%%%%%%%%%%%%%%%%%%%
We start with the 3d $\mathcal{N}=2$ $Spin(8)$ gauge theories with $N_v$ vectors, $N_s$ spinors and $N_{c}$ conjugate spinors. The corresponding 4d theories were studied in \cite{Pouliot:1995sk, Cho:1997kr}. There are three $\mathbf{8}$ dimensional representations in a $Spin(8)$ group, which are denoted as $\mathbf{8}_v, \mathbf{8}_s$ and $\mathbf{8}_c$. Those are related by triality, outer automorphism of the $D_4$ Dynkin diagram. For the purpose of listing up all the s-confinement phases, it is sufficient to consider the six cases which will be discussed in the following subsections.

When the Coulomb branch $Y$ obtains a non-zero expectation value, the gauge group is broken as
\begin{align}
so(8) & \rightarrow so(6) \times u(1) \\
\mathbf{8}_v & \rightarrow \mathbf{6}_{0} +\mathbf{1}_{2} +\mathbf{1}_{-2}\\
\mathbf{8}_s &  \rightarrow \mathbf{4}_{1}+ \overline{\mathbf{4}}_{-1} \\
\mathbf{8}_c & \rightarrow  \mathbf{4}_{-1} +  \overline{\mathbf{4}}_{1}.
\end{align}
All the components of the spinor matters are charged under the unbroken $U(1)$ gauge subgroup. Hence, they are all massive and integrated out from the low-energy spectrum. In order to obtain a stable SUSY vacuum along the Y direction, the low-energy $SO(6)$ theory also must have a stable SUSY vacuum. This is possible only for $N_v \ge 4$ \cite{Aharony:2011ci}. Therefore, the $Spin(8)$ theory only with spinor matters does not need this operator.

The second Coulomb branch $Z$ corresponds to the breaking
\begin{align}
so(8) & \rightarrow so(4) \times su(2) \times u(1) \\
\mathbf{8}_v & \rightarrow (\mathbf{4}, \mathbf{1})_{0}+(\mathbf{1}, \mathbf{2})_{ \pm 1} \\
\mathbf{8}_s &  \rightarrow (\mathbf{2}, \mathbf{2})_{0} +(\mathbf{2}^*, \mathbf{1})_{ \pm 1}\\
\mathbf{8}_c & \rightarrow  (\mathbf{2}, \mathbf{1})_{\pm 1}+(\mathbf{2}^*, \mathbf{2})_{0}.
\end{align}
Notice that the vector representation does not contain any massless field charged under the $SU(2)$ subgroup and cannot make the $SU(2)$ vacuum of the low-energy theory stable. Therefore, this branch exists only for the theory with spinor matters. When there is only a single spinor, the low-energy $SU(2)$ theory has a deformed moduli space and the origin of the moduli space is excluded from the quantum moduli space. In order that the $Z$-branch with all the matter fields turned off can  be a flat direction, the theory has to contain at least two spinors.

%%%%%%%%%%%%%%%%%%%%%%%%%%%%%%%%%%%%%%%%%%%%%%%%%%%%%%%%%%
\subsection{$(N_v,N_s,N_c)=(5,1,0)$}
%%%%%%%%%%%%%%%%%%%%%%%%%%%%%%%%%%%%%%%%%%%%%%%%%%%%%%%%%%
The first example is the 3d $\mathcal{N}=2$ $Spin(8)$ gauge theory with five vectors and one spinor. In this case, the $Y$-branch is allowed since the low-energy theory contains a 3d $\mathcal{N}=2$ $SO(6)$ gauge theory with five vectors, which has a supersymmetric vacuum and we can safely take the low-energy limit at this point. On the other hand, the $Z$-branch, where $\braket{Z}$ acquires a vev and all the matter fields are turned off, is not allowed. Consequently, we expect that there is only a single Coulomb branch parametrized by $Y$.

\begin{table}[H]\caption{3d $\mathcal{N}=2$ $Spin(8)$ theory with $(N_v,N_s,N_c)=(5,1,0)$} 
\begin{center}
\scalebox{0.85}{
  \begin{tabular}{|c||c||c|c|c|c| } \hline
  &$Spin(8)$&$SU(5)$&$U(1)_v$&$U(1)_s$&$U(1)_R$ \\ \hline
Q& $\mathbf{8}_v$&${\tiny \yng(1)}$&1&0& $R_v$ \\
$S$ & $\mathbf{8}_s$&1&0&1& $R_s$ \\ \hline
%$S'$ & $\mathbf{8}_c$&1&0&0&1& $R_c$ \\ \hline
%$\lambda$ &$\mathbf{Adj.}$&1&1&0&0&$1$  \\ \hline
$\eta=\Lambda_{N_v,N_s,N_c}^b$&1&1&$10$&$2$&$10(R_v-1)+2(R_s-1) +12=10R_v+2R_s$  \\ \hline 
$M_{QQ}:=QQ$&1&$\tiny \yng(2)$&2&0&$2R_v$ \\
$M_{SS}:=SS$&1&1&0&2&$2R_s$ \\
%$M_{S'S'}:=S'S'$&1&1&0&0&2&$2R_c$ \\
%$P_1:= SQS'$&1&${\tiny \yng(1)}$&1&1&1&$R_v+R_s+R_c$ \\
%$P_3:=SQ^3S'$&1&${\tiny \overline{\yng(1)}}$&3&1&1&$3R_v+R_s+R_c$ \\
$P_4:=SQ^4S$&1&${\tiny \overline{\yng(1)}}$&4&2&$4R_v+2R_s$ \\ \hline
%$P'_4:=S'Q^4S'$&1&1&4&0&2&$4R_v+2R_c$ \\ \hline
%$\det \, M_{QQ}$&1&1&10&0&$10R_v$  \\ \hline
%$M_{QQ}P_3^2$&1&1&8&2&2&$8R_v+2R_s+2R_c$  \\
%$M_{QQ}^3 P_1^2$&1&1&8&2&2&$8R_v+2R_s+2R_c$ \\
%$P_1P_3$&1&1&4&2&2&$4R_v+2R_s+2R_c$  \\ \hline
%$B_Q:=Q^7$&1&${\tiny \yng(1,1,1,1,1,1,1)}$&1&7&0& $7R_f$\\ 
%$B_S:=S^4$&1&1&${\tiny \yng(1,1,1,1)}$&0&$4$&$4R_S$  \\[10pt] 
%$B'_S:=S^4 Q$&1&${\tiny \yng(1)}$&${\tiny \yng(1,1,1,1)}$&1&4& $R_f+4R_S$ \\
%$P:=SQ^3S$&1&${\tiny \yng(1,1,1)}$&${\tiny \yng(2)}$&3&2& $3R_f+2R_S$ \\[9pt]
%$R:=SQ^4S$&1&${\tiny \yng(1,1,1,1)}$&${\tiny \yng(2)}$&4&2& $4R_f+2R_S$\\
%$Z:=Y_1Y_2^2Y_3 Y_4$&1&1&$-10$&$-2$&$-10-10(R_v-1)-2(R_s-1)=2-10R_v-2R_s$ \\ 
$Y:=Y_1^2 Y^2_2 Y_3 Y_4$ &1&1&$-10$&$-4$&$-12-10(R_v-1)-4(R_s-1)=2-10R_v-4R_s$ \\ \hline
%$\det  \, M_{QQ}$&1&1&1&$2N_f$&0&$2N_fR_f$  \\
%$\det \, M_{SS}$ &1&1&1&0&$2N_S$&$2N_SR_S$  \\
%&&&&&&  \\
%&&&&&&  \\
%$\eta$&1&1&1&$2N_f$&$2N_S$&$2N_f(R_f-1)+2N_S(R_S-1) +10$  \\ \hline
%$Y_1$&1&1&1&0&$-N_S(1+\mathrm{sign}(\phi_1-\phi_3))$& $-2-N_S(R_S-1)(1+\mathrm{sign}(\phi_1-\phi_2))$\\
%$Y_2$&1&1&1&0&0&$-2$\\
%$Y_3$&1&1&1&$-2N_f$&$-N_S(1-\mathrm{sign}(\phi_1-\phi_3))$& $-2-N_S(R_S-1)(1-\mathrm{sign}(\phi_1-\phi_2))  -2N_f(R_f-1) $\\
%$Z:=Y_1Y_2^2Y_3$&1&1&1&$-2N_f$&$-2N_S$& $-8 -2N_S (R_S-1) -2N_f (R_f-1)$  \\ 
%$Y:=\sqrt{Y_1 Z}$ for $\phi_1 \ge \phi_3$ &1&1&1&$-N_f$&$-2N_S$& $-5-N_f(R_f-1) -2N_S(R_S-1)$  \\
%$Z':=Y_1Y_3$&1&1&1&$-2N_f$&$-2N_S$&$-4 -2N_S (R_S-1) -2N_f (R_f-1)$ \\
%$Z'':=Y_1Y_2Y_3$&1&1&1&$-2N_f$&$-2N_S$& $-6-2N_S (R_S-1) -2N_f (R_f-1)$  \\
%$Y_{spin}:=Y_1^2Y_2^2Y_3$&1&1&1&$-2N_f$&$-4N_S$&$-10-2N_f(R_f-1)-4N_S(R_S-1)$ \\
  \end{tabular}}
  \end{center}\label{Spin8510}
\end{table}
%We find an identification between the Coulomb branch operators: $Z \sim \tilde{Y}M_{SS}$. 

The low-energy dynamics is described by $M_{QQ}, M_{SS}, P_4$ and $Y$.
The confining superpotential is constrained by the symmetries listed in Table \ref{Spin8510} and we find
\begin{align}
W=Y \left[ M_{SS}^2 \det \,  M_{QQ}+ P_4^2 M_{QQ} \right] +\eta Y M_{SS},
\end{align}
where the last term appears when we put the 4d $Spin(8)$ theory on $\mathbb{S}_1 \times \mathbb{R}_3$. $\eta$ is a dynamical scale of the 4d gauge interaction.
By integrating out the Coulomb branch operator, we can go up to the 4d $\mathcal{N}=1$ $Spin(8)$ theory with five vectors and one spinor and reproduce a deformed moduli space \cite{Pouliot:1995sk}.

%%%%%%%%%%%%%%%%%%%%%%%%%%%%%%%%%%%%%%%%%%%%%%%%%%%%%%%%%%
\subsection{$(N_v,N_s,N_c)=(4,2,0)$}
%%%%%%%%%%%%%%%%%%%%%%%%%%%%%%%%%%%%%%%%%%%%%%%%%%%%%%%%%%
The second example is the 3d $\mathcal{N}=2$ $Spin(8)$ gauge theory with four vectors and two spinors.
As in the previous case, the low-energy $SO(6)$ dynamics along the $Y$-direction is made stable by four vector matters. Along the $Z$-direction, the low-energy theory includes the $SU(2)$ gauge theory with four fundamentals and there is a stable SUSY vacuum. Therefore, we need introduce the two Coulomb branch coordinates, $Y$ and $Z$.

The low-energy dynamics is described by the Higgs branch operators $M_{QQ}, M_{SS}, P_2, P_4$ defined in Table \ref{Spin8420} and the two Coulomb branch coordinates. The confining superpotential becomes
\begin{align}
W=Z \left[ M_{SS}^2\det M_{QQ} +M_{QQ}^2P_2^2 +P_4^2\right] +Y P_2^2 ,
\end{align}
which is consistent with all the symmetries in Table \ref{Spin8420}.

\begin{table}[H]\caption{3d $\mathcal{N}=2$ $Spin(8)$ theory with $(N_v,N_s,N_c)=(4,2,0)$} 
\begin{center}
\scalebox{0.79}{
  \begin{tabular}{|c||c||c|c|c|c|c| } \hline
  &$Spin(8)$&$SU(4)$&$SU(2)$&$U(1)_v$&$U(1)_s$&$U(1)_R$ \\ \hline
Q& $\mathbf{8}_v$&${\tiny \yng(1)}$&1&0&0& $R_v$ \\
$S$ & $\mathbf{8}_s$&1&${\tiny \yng(1)}$&0&1& $R_s$ \\ \hline
%$S'$ & $\mathbf{8}_c$&1&0&0&1& $R_c$ \\ \hline
%$\lambda$ &$\mathbf{Adj.}$&1&1&0&0&$1$  \\ \hline
%$\eta=\Lambda_{N_v,N_s,N_c}^b$&1&1&$1$&$8$&4&$8(R_v-1)+4(R_s-1) +12=8R_v+4R_s$  \\ \hline 
$M_{QQ}:=QQ$&1&$\tiny \yng(2)$&1&2&0&$2R_v$ \\
$M_{SS}:=SS$&1&1&$\tiny \yng(2)$&0&2&$2R_s$ \\
%$M_{S'S'}:=S'S'$&1&1&0&0&2&$2R_c$ \\
%$P_1:= SQS'$&1&${\tiny \yng(1)}$&1&1&1&$R_v+R_s+R_c$ \\
%$P_3:=SQ^3S'$&1&${\tiny \overline{\yng(1)}}$&3&1&1&$3R_v+R_s+R_c$ \\
$P_2:=SQ^2S$&1&$\tiny \yng(1,1)$&1&2&2&$2R_v+2R_s$ \\
$P_4:=SQ^4S$&1&1&$\tiny \yng(2)$&4&2&$4R_v+2R_s$ \\ \hline
%$P'_4:=S'Q^4S'$&1&1&4&0&2&$4R_v+2R_c$ \\ \hline
%$\det \, M_{QQ}$&1&1&1&8&0&$8R_v$  \\
%$M_{SS}^2$&1&1&1&0&4&$4R_s$ \\
%$P_2^2$&1&1&1&4&4&$4R_v+4R_s$ \\
%$M_{QQ}^2P_2^2 $&1&1&1&8&4&$8R_v+4R_s$  \\ \hline
%$M_{QQ}P_3^2$&1&1&1&8&2&$8R_v+2R_s+2R_c$  \\
%$M_{QQ}^3 P_1^2$&1&1&8&2&2&$8R_v+2R_s+2R_c$ \\
%$P_1P_3$&1&1&4&2&2&$4R_v+2R_s+2R_c$  \\ \hline
%$B_Q:=Q^7$&1&${\tiny \yng(1,1,1,1,1,1,1)}$&1&7&0& $7R_f$\\ 
%$B_S:=S^4$&1&1&${\tiny \yng(1,1,1,1)}$&0&$4$&$4R_S$  \\[10pt] 
%$B'_S:=S^4 Q$&1&${\tiny \yng(1)}$&${\tiny \yng(1,1,1,1)}$&1&4& $R_f+4R_S$ \\
%$P:=SQ^3S$&1&${\tiny \yng(1,1,1)}$&${\tiny \yng(2)}$&3&2& $3R_f+2R_S$ \\[9pt]
%$R:=SQ^4S$&1&${\tiny \yng(1,1,1,1)}$&${\tiny \yng(2)}$&4&2& $4R_f+2R_S$\\
$Z:=Y_1Y_2^2Y_3 Y_4$&1&1&$1$&$-8$&$-4$&$-10-8(R_v-1)-4(R_s-1)=2-8R_v-4R_s$ \\ 
$Y:=\sqrt{Y_1^2Y_2^2Y_3 Y_4}$&1&1&$1$&$-4$&$-4$&$-6-4(R_v-1)-4(R_s-1)=2-4R_v-4R_s$ \\ \hline
%$\det  \, M_{QQ}$&1&1&1&$2N_f$&0&$2N_fR_f$  \\
%$\det \, M_{SS}$ &1&1&1&0&$2N_S$&$2N_SR_S$  \\
%&&&&&&  \\
%&&&&&&  \\
%$\eta$&1&1&1&$2N_f$&$2N_S$&$2N_f(R_f-1)+2N_S(R_S-1) +10$  \\ \hline
%$Y_1$&1&1&1&0&$-N_S(1+\mathrm{sign}(\phi_1-\phi_3))$& $-2-N_S(R_S-1)(1+\mathrm{sign}(\phi_1-\phi_2))$\\
%$Y_2$&1&1&1&0&0&$-2$\\
%$Y_3$&1&1&1&$-2N_f$&$-N_S(1-\mathrm{sign}(\phi_1-\phi_3))$& $-2-N_S(R_S-1)(1-\mathrm{sign}(\phi_1-\phi_2))  -2N_f(R_f-1) $\\
%$Z:=Y_1Y_2^2Y_3$&1&1&1&$-2N_f$&$-2N_S$& $-8 -2N_S (R_S-1) -2N_f (R_f-1)$  \\ 
%$Y:=\sqrt{Y_1 Z}$ for $\phi_1 \ge \phi_3$ &1&1&1&$-N_f$&$-2N_S$& $-5-N_f(R_f-1) -2N_S(R_S-1)$  \\
%$Z':=Y_1Y_3$&1&1&1&$-2N_f$&$-2N_S$&$-4 -2N_S (R_S-1) -2N_f (R_f-1)$ \\
%$Z'':=Y_1Y_2Y_3$&1&1&1&$-2N_f$&$-2N_S$& $-6-2N_S (R_S-1) -2N_f (R_f-1)$  \\
%$Y_{spin}:=Y_1^2Y_2^2Y_3$&1&1&1&$-2N_f$&$-4N_S$&$-10-2N_f(R_f-1)-4N_S(R_S-1)$ \\
  \end{tabular}}
  \end{center}\label{Spin8420}
\end{table}

%%%%%%%%%%%%%%%%%%%%%%%%%%%%%%%%%%%%%%%%%%%%%%%%%%%%%%%%%%
\subsection{$(N_v,N_s,N_c)=(4,1,1)$}
%%%%%%%%%%%%%%%%%%%%%%%%%%%%%%%%%%%%%%%%%%%%%%%%%%%%%%%%%%
Let us study the case where we introduce both spinor and conjugate spinor matters. The s-confinement phase appears in the 3d $\mathcal{N}=2$ $Spin(8)$ gauge theory with four vectors, one spinor and one conjugate spinor. The corresponding 4d theory was studied in \cite{Cho:1997kr}. The Higgs branch is identical to the 4d case and parametrized by three mesons $M_{QQ}, M_{SS}, M_{S'S'}$ and four vector-spinor composites $P_1, P_3, P_4, P'_4$ defined in Table \ref{Spin8411}. 

The Coulomb branch $Y$ is allowed since the four vectors $\mathbf{6}_{0} \in \mathbf{8}_v $ can make this direction stable. The $Z$-direction is also allowed due to the two spinors. The matter contents and their quantum numbers are summarized in Table \ref{Spin8411} which includes the dynamical scale of the gauge interaction in the corresponding 4d $\mathcal{N}=1$ $Spin(8)$ theory. The superpotential becomes
\begin{align}
W&=Z \left[ M_{SS} M_{S'S'} \det \, M_{QQ}+M_{QQ}^3P_1^2+P_3M_{QQ}P_3+P_4P'_4\right] \nonumber \\
&\qquad + Y \left[ P_1P_3 +M_{SS}P'_4+M_{S'S'}P_4 \right] +\eta Z,
\end{align}
where the last term appears only when we put the 4d theory on $\mathbb{S}^1 \times \mathbb{R}^3$. By integrating out the Coulomb branch operators, we can reproduce the deformed and un-deformed constraints of the 4d theory \cite{Cho:1997kr}.

\begin{table}[H]\caption{Quantum numbers for 3d $\mathcal{N}=2$ $Spin(8)$ theory with $(N_v,N_s,N_c)=(4,1,1)$} 
\begin{center}
\scalebox{0.72}{
  \begin{tabular}{|c||c||c|c|c|c|c| } \hline
  &$Spin(8)$&$SU(4)$&$U(1)_v$&$U(1)_s$&$U(1)_c$&$U(1)_R$ \\ \hline
Q& $\mathbf{8}_v$&${\tiny \yng(1)}$&1&0&0& $R_v$ \\
$S$ & $\mathbf{8}_s$&1&0&1&0& $R_s$ \\
$S'$ & $\mathbf{8}_c$&1&0&0&1& $R_c$ \\ \hline
%$\lambda$ &$\mathbf{Adj.}$&1&1&0&0&$1$  \\ \hline
$\eta=\Lambda_{N_v,N_s,N_c}^b$&1&1&$8$&$2$&2&$8(R_v-1)+2(R_s-1)+2(R_c-1) +12=8R_v+2R_s+2R_c$  \\ \hline 
$M_{QQ}:=QQ$&1&$\tiny \yng(2)$&2&0&0&$2R_v$ \\
$M_{SS}:=SS$&1&1&0&2&0&$2R_s$ \\
$M_{S'S'}:=S'S'$&1&1&0&0&2&$2R_c$ \\
$P_1:= SQS'$&1&${\tiny \yng(1)}$&1&1&1&$R_v+R_s+R_c$ \\
$P_3:=SQ^3S'$&1&${\tiny \overline{\yng(1)}}$&3&1&1&$3R_v+R_s+R_c$ \\
$P_4:=SQ^4S$&1&1&4&2&0&$4R_v+2R_s$ \\
$P'_4:=S'Q^4S'$&1&1&4&0&2&$4R_v+2R_c$ \\ \hline
$\det \, M_{QQ}$&1&1&8&0&0&$8R_v$  \\
$M_{QQ}P_3^2$&1&1&8&2&2&$8R_v+2R_s+2R_c$  \\
$M_{QQ}^3 P_1^2$&1&1&8&2&2&$8R_v+2R_s+2R_c$ \\
$P_1P_3$&1&1&4&2&2&$4R_v+2R_s+2R_c$  \\ \hline
%$B_Q:=Q^7$&1&${\tiny \yng(1,1,1,1,1,1,1)}$&1&7&0& $7R_f$\\ 
%$B_S:=S^4$&1&1&${\tiny \yng(1,1,1,1)}$&0&$4$&$4R_S$  \\[10pt] 
%$B'_S:=S^4 Q$&1&${\tiny \yng(1)}$&${\tiny \yng(1,1,1,1)}$&1&4& $R_f+4R_S$ \\
%$P:=SQ^3S$&1&${\tiny \yng(1,1,1)}$&${\tiny \yng(2)}$&3&2& $3R_f+2R_S$ \\[9pt]
%$R:=SQ^4S$&1&${\tiny \yng(1,1,1,1)}$&${\tiny \yng(2)}$&4&2& $4R_f+2R_S$\\
$Z:=Y_1Y_2^2Y_3 Y_4$&1&1&$-8$&$-2$&$-2$&$2-8R_v-2R_s-2R_c$ \\ 
$Y:=\sqrt{Y_1^2Y_2^2Y_3 Y_4}$&1&1&$-4$&$-2$&$-2$&$2-4R_v-2R_s-2R_c$ \\ \hline
%$\det  \, M_{QQ}$&1&1&1&$2N_f$&0&$2N_fR_f$  \\
%$\det \, M_{SS}$ &1&1&1&0&$2N_S$&$2N_SR_S$  \\
%&&&&&&  \\
%&&&&&&  \\
%$\eta$&1&1&1&$2N_f$&$2N_S$&$2N_f(R_f-1)+2N_S(R_S-1) +10$  \\ \hline
%$Y_1$&1&1&1&0&$-N_S(1+\mathrm{sign}(\phi_1-\phi_3))$& $-2-N_S(R_S-1)(1+\mathrm{sign}(\phi_1-\phi_2))$\\
%$Y_2$&1&1&1&0&0&$-2$\\
%$Y_3$&1&1&1&$-2N_f$&$-N_S(1-\mathrm{sign}(\phi_1-\phi_3))$& $-2-N_S(R_S-1)(1-\mathrm{sign}(\phi_1-\phi_2))  -2N_f(R_f-1) $\\
%$Z:=Y_1Y_2^2Y_3$&1&1&1&$-2N_f$&$-2N_S$& $-8 -2N_S (R_S-1) -2N_f (R_f-1)$  \\ 
%$Y:=\sqrt{Y_1 Z}$ for $\phi_1 \ge \phi_3$ &1&1&1&$-N_f$&$-2N_S$& $-5-N_f(R_f-1) -2N_S(R_S-1)$  \\
%$Z':=Y_1Y_3$&1&1&1&$-2N_f$&$-2N_S$&$-4 -2N_S (R_S-1) -2N_f (R_f-1)$ \\
%$Z'':=Y_1Y_2Y_3$&1&1&1&$-2N_f$&$-2N_S$& $-6-2N_S (R_S-1) -2N_f (R_f-1)$  \\
%$Y_{spin}:=Y_1^2Y_2^2Y_3$&1&1&1&$-2N_f$&$-4N_S$&$-10-2N_f(R_f-1)-4N_S(R_S-1)$ \\
  \end{tabular}}
  \end{center}\label{Spin8411}
\end{table}

%%%%%%%%%%%%%%%%%%%%%%%%%%%%%%%%%%%%%%%%%%%%%%%%%%%%%%%%%%
\subsection{$(N_v,N_s,N_c)=(3,3,0)$}
%%%%%%%%%%%%%%%%%%%%%%%%%%%%%%%%%%%%%%%%%%%%%%%%%%%%%%%%%%
Let us consider the 3d $\mathcal{N}=2$ $Spin(8)$ gauge theory with three vectors and three spinors. 
The $Y$-branch is not allowed since the low-energy $SO(6)$ gauge theory contains only three vectors and there is no stable SUSY vacuum. Along the $Z$-branch, the resulting $SO(4) \times SU(2)$ gauge theory obtains a stable SUSY vacuum due to the sufficient number of matter fields to stabilize the vacuum. 

The low-energy dynamics is described by the four chiral superfields $M_{QQ}, M_{SS}, P_2$ and $Z$, which are defined in Table \ref{Spin8330}. By using the symmetries listed in Table \ref{Spin8330}, the confining superpotential is determined as
\begin{align}
W=Z \left[ \det M_{QQ}\det M_{SS}  +M_{QQ} M_{SS} P_2^2 \right] .
\end{align}
%
%The last term appears when we put the $Spin(8)$ theory on $\mathbb{S}_1 \times \mathbb{R}_3$.

\begin{table}[H]\caption{Quantum numbers for 3d $\mathcal{N}=2$ $Spin(8)$ theory with $(N_v,N_s,N_c)=(3,3,0)$} 
\begin{center}
\scalebox{0.81}{
  \begin{tabular}{|c||c||c|c|c|c|c| } \hline
  &$Spin(8)$&$SU(3)$&$SU(3)$&$U(1)_v$&$U(1)_s$&$U(1)_R$ \\ \hline
Q& $\mathbf{8}_v$&${\tiny \yng(1)}$&1&1&0& $R_v$ \\
$S$ & $\mathbf{8}_s$&1&${\tiny \yng(1)}$&0&1& $R_s$ \\ \hline
%$S'$ & $\mathbf{8}_c$&1&0&0&1& $R_c$ \\ \hline
%$\lambda$ &$\mathbf{Adj.}$&1&1&0&0&$1$  \\ \hline
%$\eta=\Lambda_{N_v,N_s,N_c}^b$&1&1&$1$&$6$&6&$6(R_v-1)+6(R_s-1) +12=6R_v+6R_s$  \\ \hline 
$M_{QQ}:=QQ$&1&$\tiny \yng(2)$&1&2&0&$2R_v$ \\
$M_{SS}:=SS$&1&1&$\tiny \yng(2)$&0&2&$2R_s$ \\
%$M_{S'S'}:=S'S'$&1&1&0&0&2&$2R_c$ \\
%$P_1:= SQS'$&1&${\tiny \yng(1)}$&1&1&1&$R_v+R_s+R_c$ \\
%$P_3:=SQ^3S'$&1&${\tiny \overline{\yng(1)}}$&3&1&1&$3R_v+R_s+R_c$ \\
$P_2:=SQ^2S$&1&$\tiny \overline{\yng(1)}$&$\tiny \overline{\yng(1)}$&2&2&$2R_v+2R_s$ \\ \hline
%$P_4:=SQ^4S$&1&1&4&2&0&$4R_v+2R_s$ \\ \hline
%$P'_4:=S'Q^4S'$&1&1&4&0&2&$4R_v+2R_c$ \\ \hline
%$\det \, M_{QQ}$&1&1&1&6&0&$6R_v$  \\
%$\det \, M_{SS}$&1&1&1&0&6&$6R_s$ \\
%$P_2^2$&1&1&1&4&4&$4R_v+4R_s$ \\
%$M_{QQ}P_2^2 $&1&1&1&6&4&$6R_v+4R_s$  \\ \hline
%$M_{QQ}P_3^2$&1&1&1&8&2&$8R_v+2R_s+2R_c$  \\
%$M_{QQ}^3 P_1^2$&1&1&8&2&2&$8R_v+2R_s+2R_c$ \\
%$P_1P_3$&1&1&4&2&2&$4R_v+2R_s+2R_c$  \\ \hline
%$B_Q:=Q^7$&1&${\tiny \yng(1,1,1,1,1,1,1)}$&1&7&0& $7R_f$\\ 
%$B_S:=S^4$&1&1&${\tiny \yng(1,1,1,1)}$&0&$4$&$4R_S$  \\[10pt] 
%$B'_S:=S^4 Q$&1&${\tiny \yng(1)}$&${\tiny \yng(1,1,1,1)}$&1&4& $R_f+4R_S$ \\
%$P:=SQ^3S$&1&${\tiny \yng(1,1,1)}$&${\tiny \yng(2)}$&3&2& $3R_f+2R_S$ \\[9pt]
%$R:=SQ^4S$&1&${\tiny \yng(1,1,1,1)}$&${\tiny \yng(2)}$&4&2& $4R_f+2R_S$\\
$Z:=Y_1Y_2^2Y_3 Y_4$&1&1&$1$&$-6$&$-6$&$-10-6(R_v-1)-6(R_s-1)=2-6R_v-6R_s$ \\ \hline
%$Y:=\sqrt{Y_1Z}$ for $\phi_1> \phi_3, \phi_1> \phi_4$&1&1&$1$&$-3$&$-6$&$-6-3(R_v-1)-6(R_s-1)=3-3R_v-6R_s$ \\ \hline
%$\det  \, M_{QQ}$&1&1&1&$2N_f$&0&$2N_fR_f$  \\
%$\det \, M_{SS}$ &1&1&1&0&$2N_S$&$2N_SR_S$  \\
%&&&&&&  \\
%&&&&&&  \\
%$\eta$&1&1&1&$2N_f$&$2N_S$&$2N_f(R_f-1)+2N_S(R_S-1) +10$  \\ \hline
%$Y_1$&1&1&1&0&$-N_S(1+\mathrm{sign}(\phi_1-\phi_3))$& $-2-N_S(R_S-1)(1+\mathrm{sign}(\phi_1-\phi_2))$\\
%$Y_2$&1&1&1&0&0&$-2$\\
%$Y_3$&1&1&1&$-2N_f$&$-N_S(1-\mathrm{sign}(\phi_1-\phi_3))$& $-2-N_S(R_S-1)(1-\mathrm{sign}(\phi_1-\phi_2))  -2N_f(R_f-1) $\\
%$Z:=Y_1Y_2^2Y_3$&1&1&1&$-2N_f$&$-2N_S$& $-8 -2N_S (R_S-1) -2N_f (R_f-1)$  \\ 
%$Y:=\sqrt{Y_1 Z}$ for $\phi_1 \ge \phi_3$ &1&1&1&$-N_f$&$-2N_S$& $-5-N_f(R_f-1) -2N_S(R_S-1)$  \\
%$Z':=Y_1Y_3$&1&1&1&$-2N_f$&$-2N_S$&$-4 -2N_S (R_S-1) -2N_f (R_f-1)$ \\
%$Z'':=Y_1Y_2Y_3$&1&1&1&$-2N_f$&$-2N_S$& $-6-2N_S (R_S-1) -2N_f (R_f-1)$  \\
%$Y_{spin}:=Y_1^2Y_2^2Y_3$&1&1&1&$-2N_f$&$-4N_S$&$-10-2N_f(R_f-1)-4N_S(R_S-1)$ \\
  \end{tabular}}
  \end{center}\label{Spin8330}
\end{table}

%%%%%%%%%%%%%%%%%%%%%%%%%%%%%%%%%%%%%%%%%%%%%%%%%%%%%%%%%%
\subsection{$(N_v,N_s,N_c)=(3,2,1)$}
%%%%%%%%%%%%%%%%%%%%%%%%%%%%%%%%%%%%%%%%%%%%%%%%%%%%%%%%%%
The next example is the 3d $\mathcal{N}=2$ $Spin(8)$ gauge theory with three vectors, two spinors and one conjugate spinor. The analysis of the Coulomb branch is the same as the previous example. Since the number of the vector matters is less than four, the $Y$-branch cannot be a stable vacuum. Along the $Z$-direction, there are plenty of matter fields charged under the $so(4) \times su(2)$ and the $Z$-direction can be made stable and supersymmetric.

The Higgs branch is described by the six composite operators, $M_{QQ}, M_{SS}, M_{S'S'}, P_1, P_2 $ and $P_3$, which are defined in Table \ref{Spin8321}.  Table \ref{Spin8321} summarizes the quantum numbers of the moduli coordinates. The confining superpotential takes 
\begin{align}
W=Z \left[ \det M_{QQ} \det M_{SS} M_{S'S'} +P_1P_2P_3+M_{QQ}P_2^2M_{S'S'}+M_{SS}P_3^2+M_{QQ}^2M_{SS}P_1^2  \right].
\end{align}
\begin{table}[H]\caption{3d $\mathcal{N}=2$ $Spin(8)$ theory with $(N_v,N_s,N_c)=(3,2,1)$} 
\begin{center}
\scalebox{0.93}{
  \begin{tabular}{|c||c||c|c|c|c|c|c| } \hline
  &$Spin(8)$&$SU(3)$&$SU(2)$&$U(1)_v$&$U(1)_s$&$U(1)_c$&$U(1)_R$ \\ \hline
Q& $\mathbf{8}_v$&${\tiny \yng(1)}$&1&1&0&0& $R_v$ \\
$S$ & $\mathbf{8}_s$&1&${\tiny \yng(1)}$&0&1&0& $R_s$ \\ 
$S'$ & $\mathbf{8}_c$&1&1&0&0&1& $R_c$ \\ \hline
%$\eta=\Lambda_{N_v,N_s,N_c}^b$&1&1&$1$&$6$&4&2&$6(R_v-1)+4(R_s-1)+2(R_c-1) +12=6R_v+4R_s+2R_c$  \\ \hline 
$M_{QQ}:=QQ$&1&$\tiny \yng(2)$&1&2&0&0&$2R_v$ \\
$M_{SS}:=SS$&1&1&$\tiny \yng(2)$&0&2&0&$2R_s$ \\
$M_{S'S'}:=S'S'$&1&1&1&0&0&2&$2R_c$ \\
$P_1:= SQS'$&1&${\tiny \yng(1)}$&${\tiny \yng(1)}$&1&1&1&$R_v+R_s+R_c$ \\
$P_2:=SQ^2S$&1&$\tiny \overline{\yng(1)}$&1&2&2&0&$2R_v+2R_s$ \\ 
$P_3:=SQ^3S'$&1&1&${\tiny \yng(1)}$&3&1&1&$3R_v+R_s+R_c$ \\ \hline
%$P_2:=SQ^2S$&1&$\tiny \overline{\yng(1)}$&1&2&2&$2R_v+2R_s$ \\ \hline
%$P_4:=SQ^4S$&1&1&4&2&0&$4R_v+2R_s$ \\ \hline
%$P'_4:=S'Q^4S'$&1&1&4&0&2&$4R_v+2R_c$ \\ \hline
%$\det \, M_{QQ}$&1&1&1&6&0&0&$6R_v$  \\
%$\det \, M_{SS}$&1&1&1&0&4&0&$4R_s$ \\
%$M_{S'S'}$&1&1&1&0&0&2&$2R_c$ \\
%$P_1P_2P_3$&1&1&1&6&4&2&$6R_v+4R_s+2R_c$ \\
%$P_2^2$&1&1&1&4&4&$4R_v+4R_s$ \\
%$M_{QQ}P_2^2 $&1&1&1&6&4&0&$6R_v+4R_s$  \\ 
%$M_{SS}P_3^2$&1&1&1&6&4&2&$6R_v+4R_s+2R_c$\\
%$M_{QQ}^2M_{SS}P_1^2 $&1&1&1&6&4&2& $6R_v+4R_s+2R_c$\\ \hline
%$M_{QQ}P_3^2$&1&1&1&8&2&$8R_v+2R_s+2R_c$  \\
%$M_{QQ}^3 P_1^2$&1&1&8&2&2&$8R_v+2R_s+2R_c$ \\
%$P_1P_3$&1&1&4&2&2&$4R_v+2R_s+2R_c$  \\ \hline
%$B_Q:=Q^7$&1&${\tiny \yng(1,1,1,1,1,1,1)}$&1&7&0& $7R_f$\\ 
%$B_S:=S^4$&1&1&${\tiny \yng(1,1,1,1)}$&0&$4$&$4R_S$  \\[10pt] 
%$B'_S:=S^4 Q$&1&${\tiny \yng(1)}$&${\tiny \yng(1,1,1,1)}$&1&4& $R_f+4R_S$ \\
%$P:=SQ^3S$&1&${\tiny \yng(1,1,1)}$&${\tiny \yng(2)}$&3&2& $3R_f+2R_S$ \\[9pt]
%$R:=SQ^4S$&1&${\tiny \yng(1,1,1,1)}$&${\tiny \yng(2)}$&4&2& $4R_f+2R_S$\\
$Z:=Y_1Y_2^2Y_3Y_4$&1&1&$1$&$-6$&$-4$&$-2$&$2-6R_v-4R_s-2R_c$ \\ \hline
%$Y:=\sqrt{Y_1Z}$ for $\phi_1> \phi_3, \phi_1> \phi_4$&1&1&$1$&$-3$&$-6$&$-6-3(R_v-1)-6(R_s-1)=3-3R_v-6R_s$ \\ \hline
%$\det  \, M_{QQ}$&1&1&1&$2N_f$&0&$2N_fR_f$  \\
%$\det \, M_{SS}$ &1&1&1&0&$2N_S$&$2N_SR_S$  \\
%&&&&&&  \\
%&&&&&&  \\
%$\eta$&1&1&1&$2N_f$&$2N_S$&$2N_f(R_f-1)+2N_S(R_S-1) +10$  \\ \hline
%$Y_1$&1&1&1&0&$-N_S(1+\mathrm{sign}(\phi_1-\phi_3))$& $-2-N_S(R_S-1)(1+\mathrm{sign}(\phi_1-\phi_2))$\\
%$Y_2$&1&1&1&0&0&$-2$\\
%$Y_3$&1&1&1&$-2N_f$&$-N_S(1-\mathrm{sign}(\phi_1-\phi_3))$& $-2-N_S(R_S-1)(1-\mathrm{sign}(\phi_1-\phi_2))  -2N_f(R_f-1) $\\
%$Z:=Y_1Y_2^2Y_3$&1&1&1&$-2N_f$&$-2N_S$& $-8 -2N_S (R_S-1) -2N_f (R_f-1)$  \\ 
%$Y:=\sqrt{Y_1 Z}$ for $\phi_1 \ge \phi_3$ &1&1&1&$-N_f$&$-2N_S$& $-5-N_f(R_f-1) -2N_S(R_S-1)$  \\
%$Z':=Y_1Y_3$&1&1&1&$-2N_f$&$-2N_S$&$-4 -2N_S (R_S-1) -2N_f (R_f-1)$ \\
%$Z'':=Y_1Y_2Y_3$&1&1&1&$-2N_f$&$-2N_S$& $-6-2N_S (R_S-1) -2N_f (R_f-1)$  \\
%$Y_{spin}:=Y_1^2Y_2^2Y_3$&1&1&1&$-2N_f$&$-4N_S$&$-10-2N_f(R_f-1)-4N_S(R_S-1)$ \\
  \end{tabular}}
  \end{center}\label{Spin8321}
\end{table}

%%%%%%%%%%%%%%%%%%%%%%%%%%%%%%%%%%%%%%%%%%%%%%%%%%%%%%%%%%
\subsection{$(N_v,N_s,N_c)=(2,2,2)$}
%%%%%%%%%%%%%%%%%%%%%%%%%%%%%%%%%%%%%%%%%%%%%%%%%%%%%%%%%%
The final example of the $Spin(8)$ s-confinement phases is the 3d $\mathcal{N}=2$ $Spin(8)$ gauge theory with two vectors, two spinors and two conjugate spinors. The theory has a one-dimensional Coulomb branch labeled by $Z$. The $Y$-branch is excluded from the moduli space of vacua since the low-energy $SO(6)$ theory along this direction does not have enough vector matters to realize the stable supersymmetric vacuum and the runaway superpotential is generated. 

The low-energy dynamics is described by $M_{QQ}, M_{SS}, M_{S'S'}, P_1, P_2, P'_2, B, F$ and $Z$ whose quantum numbers are summarized in Table \ref{Spin8222}. The confining superpotential becomes
\begin{align}
W= Z \left[ M_{QQ}^2M_{SS}^2M_{S'S'}^2 +M_{QQ}^2B^2+M_{SS}^2{P'_2}^2 +M_{S'S'}^2P_2^2 +P_2 P'_2 B +F^2\right].
\end{align}

\begin{table}[H]\caption{3d $\mathcal{N}=2$ $Spin(8)$ theory with $(N_v,N_s,N_c)=(2,2,2)$} 
\begin{center}
\scalebox{0.85}{
  \begin{tabular}{|c||c||c|c|c|c|c|c|c| } \hline
  &$Spin(8)$&$SU(2)$&$SU(2)$&$SU(2)$&$U(1)_v$&$U(1)_s$&$U(1)_c$&$U(1)_R$ \\ \hline
Q& $\mathbf{8}_v$&${\tiny \yng(1)}$&1&1&1&0&0& $R_v$ \\
$S$ & $\mathbf{8}_s$&1&${\tiny \yng(1)}$&1&0&1&0& $R_s$ \\
$S'$ & $\mathbf{8}_c$&1&1&${\tiny \yng(1)}$&0&0&1& $R_c$ \\ \hline
%$\eta=\Lambda_{N_v,N_S,N_c}^b$&1&1&1&1&$4$&$4$&4&$4(R_v-1)+4(R_s-1)+4(R_c-1) +12$  \\ \hline 
$M_{QQ}:=QQ$&1&$\tiny \yng(2)$&1&1&2&0&0&$2R_v$ \\
$M_{SS}:=SS$&1&1&$\tiny \yng(2)$&1&0&2&0&$2R_s$ \\
$M_{S'S'}:=S'S'$&1&1&1&$\tiny \yng(2)$&0&0&2&$2R_c$ \\
$P_1:=SQS'$&1&${\tiny \yng(1)}$&${\tiny \yng(1)}$&${\tiny \yng(1)}$&1&1&1&$R_v+R_s+R_c$ \\
$P_2:=SQ^2S$&1&1&1&1&2&2&0&$2R_v+2R_s$ \\ 
$P'_2:=S'Q^2S'$&1&1&1&1&2&0&2&$2R_v+2R_c$ \\
$B:=S^2S'^2$&1&1&1&1&0&2&2&$2R_s+2R_c$ \\
$F:=S^2S'^2Q^2$&1&1&1&1&2&2&2&$2R_v+2R_s+2R_c$ \\ \hline
%$B_Q:=Q^7$&1&${\tiny \yng(1,1,1,1,1,1,1)}$&1&7&0& $7R_f$\\ 
%$B_S:=S^4$&1&1&${\tiny \yng(1,1,1,1)}$&0&$4$&$4R_S$  \\[10pt] 
%$B'_S:=S^4 Q$&1&${\tiny \yng(1)}$&${\tiny \yng(1,1,1,1)}$&1&4& $R_f+4R_S$ \\
%$P:=SQ^3S$&1&${\tiny \yng(1,1,1)}$&${\tiny \yng(2)}$&3&2& $3R_f+2R_S$ \\[9pt]
%$R:=SQ^4S$&1&${\tiny \yng(1,1,1,1)}$&${\tiny \yng(2)}$&4&2& $4R_f+2R_S$\\
%&&&&&&&& \\ \hline
$Z:=Y_1Y_2^2Y_3Y_4$&1&1&$1$&1&$-4$&$-4$&$-4$&$2-4R_v-4R_s-4R_c$ \\  \hline
%$\det  \, M_{QQ}$&1&1&1&$2N_f$&0&$2N_fR_f$  \\
%$\det \, M_{SS}$ &1&1&1&0&$2N_S$&$2N_SR_S$  \\
%&&&&&&  \\
%&&&&&&  \\
%$\eta$&1&1&1&$2N_f$&$2N_S$&$2N_f(R_f-1)+2N_S(R_S-1) +10$  \\ \hline
%$Y_1$&1&1&1&0&$-N_S(1+\mathrm{sign}(\phi_1-\phi_3))$& $-2-N_S(R_S-1)(1+\mathrm{sign}(\phi_1-\phi_2))$\\
%$Y_2$&1&1&1&0&0&$-2$\\
%$Y_3$&1&1&1&$-2N_f$&$-N_S(1-\mathrm{sign}(\phi_1-\phi_3))$& $-2-N_S(R_S-1)(1-\mathrm{sign}(\phi_1-\phi_2))  -2N_f(R_f-1) $\\
%$Z:=Y_1Y_2^2Y_3$&1&1&1&$-2N_f$&$-2N_S$& $-8 -2N_S (R_S-1) -2N_f (R_f-1)$  \\ 
%$Y:=\sqrt{Y_1 Z}$ for $\phi_1 \ge \phi_3$ &1&1&1&$-N_f$&$-2N_S$& $-5-N_f(R_f-1) -2N_S(R_S-1)$  \\
%$Z':=Y_1Y_3$&1&1&1&$-2N_f$&$-2N_S$&$-4 -2N_S (R_S-1) -2N_f (R_f-1)$ \\
%$Z'':=Y_1Y_2Y_3$&1&1&1&$-2N_f$&$-2N_S$& $-6-2N_S (R_S-1) -2N_f (R_f-1)$  \\
%$Y_{spin}:=Y_1^2Y_2^2Y_3$&1&1&1&$-2N_f$&$-4N_S$&$-10-2N_f(R_f-1)-4N_S(R_S-1)$ \\
  \end{tabular}}
  \end{center}\label{Spin8222}
\end{table}

%%%%%%%%%%%%%%%%%%%%%%%%%%%%%%%%%%%%%%%%%%%%%%%%%%%%%%%%%%
%%%%%%%%%%%%%%%%%%%%%%%%%%%%%%%%%%%%%%%%%%%%%%%%%%%%%%%%%%
%%%%%%%%%%%%%%%%%%%%%%%%%%%%%%%%%%%%%%%%%%%%%%%%%%%%%%%%%%
\section{$Spin(9)$ theories}
%%%%%%%%%%%%%%%%%%%%%%%%%%%%%%%%%%%%%%%%%%%%%%%%%%%%%%%%%%
%%%%%%%%%%%%%%%%%%%%%%%%%%%%%%%%%%%%%%%%%%%%%%%%%%%%%%%%%%
%%%%%%%%%%%%%%%%%%%%%%%%%%%%%%%%%%%%%%%%%%%%%%%%%%%%%%%%%%
Let us move on to the 3d $\mathcal{N}=2$ $Spin(9)$ gauge theories with $N_v$ vectors and $N_s$ spinors. When the Coulomb branch operator $Y$ obtains a non-zero vacuum expectation value, the gauge group is broken as
\begin{align}
so(9) &\rightarrow so(7) \times u(1) \\
\mathbf{9} &\rightarrow \mathbf{7}_0 +\mathbf{1} _2 +\mathbf{1} _{-2} \\
\mathbf{16} & \rightarrow \mathbf{8}_1 +\mathbf{8}_{-1}.   
\end{align}
Thus, the spinor matters are all massive and integrated out while the vector matters reduce to the massless $\mathbf{7}$ fields. For the theories only with spinors, this branch is not allowed since the low-energy $SO(7)$ pure SYM has no stable SUSY vacuum. For the theories with $N_v (\ge 5)$ vectors, on the other hand, the low-energy $SO(7)$ SQCD can have a stable SUSY vacuum at the origin of moduli space. Therefore, for $N_v \ge 5$, we need to introduce this coordinate. 

The second Coulomb branch is denoted as $Z$ and its expectation value breaks the gauge group as
\begin{align}
so(9) & \rightarrow so(5) \times su(2) \times u(1) \\
\mathbf{9} &\rightarrow (\mathbf{5},\mathbf{1})_0 +(\mathbf{1},\mathbf{2})_{\pm 1} \\
\mathbf{16} &\rightarrow  (\mathbf{4},\mathbf{2})_0 +(\mathbf{4},\mathbf{1})_{\pm 1}.
\end{align}
For the theories only with vectors, this branch is not allowed since the low-energy $SU(2)$ gauge theory has no dynamical field and its vacuum becomes runaway. When the theory includes the spinor matters, the low-energy $SO(5) \times SU(2)$ theory can obtain a stable SUSY vacuum due to the presence of $(\mathbf{4},\mathbf{2})_0$. Therefore, we need to introduce a $Z$ coordinate for the theories with spinors.

When $N_s \ge 4$, there could be an additional Coulomb branch $V$ which corresponds to the breaking
\begin{align}
so(9) & \rightarrow  su(4) \times u(1) \\
\mathbf{9} &\rightarrow  \mathbf{1}_{0} +\mathbf{4}_{1} +\mathbf{\overline{4}}_{-1} \\
\mathbf{16} &\rightarrow \mathbf{6}_{0}+ \mathbf{4}_{-1} +\mathbf{\overline{4}}_{1}  +\mathbf{1}_{2} +\mathbf{1}_{-2}.
\end{align}
Almost all the components of the vector matter are massive and reduce to a singlet. The spinor matter reduces to $\mathbf{6}_{0}$ and the dynamics of the $SO(6) \simeq SU(4)$ theory has a stable SUSY vacuum for $N_s \ge 4$. In the following subsection, we will only consider the theories with $N_s \le 3$ spinors and this operator does not appear.

%%%%%%%%%%%%%%%%%%%%%%%%%%%%%%%%%%%%%%%%%%%%%%%%%%%%%%%%%%
\subsection{$(N_v,N_s)=(5,1)$}
%%%%%%%%%%%%%%%%%%%%%%%%%%%%%%%%%%%%%%%%%%%%%%%%%%%%%%%%%%
The first example of the $Spin(9)$ s-confinement is the 3d $\mathcal{N}=2$ $Spin(9)$ gauge theory with five vectors and one spinor. In this case, we need to introduce the two Coulomb branch coordinates $Z$ and $Y$. The Higgs branch is described by the five composite operators, $M_{QQ}, M_{SS}, P_1, P_4$ and $P_5$ defined in Table \ref{Spin951}. The confining superpotential becomes
\begin{align}
W=Z\left[ M_{SS}^2 \det \, M_{QQ} +M_{QQ}^4P_1^2 +M_{QQ}P_4^2 +P_5^2 \right]+ Y \left[ P_1P_4 +M_{SS} P_5 \right] +\eta Z,
\end{align}
where the last term appears when we consider the corresponding 4d theory on a circle. By integrating out the Coulomb branches, we can reproduce the quantum-deformed moduli space of the 4d theory \cite{Cho:1997kr}.

\begin{table}[H]\caption{3d $\mathcal{N}=2$ $Spin(9)$ theory with $(N_v,N_s)=(5,1)$} 
\begin{center}
\scalebox{1}{
  \begin{tabular}{|c||c||c|c|c|c| } \hline
  &$Spin(9)$&$SU(5)$&$U(1)_v$&$U(1)_s$&$U(1)_R$ \\ \hline
Q& $\mathbf{9}$&${\tiny \yng(1)}$&1&0& $R_v$ \\
$S$ & $\mathbf{16}$&1&0&1& $R_s$ \\ \hline
%$\lambda$ &$\mathbf{Adj.}$&1&1&0&0&$1$  \\ \hline
$\eta=\Lambda_{N_v,N_s}^b$&1&1&$10$&$4$&$10R_v+4R_s$  \\ \hline 
$M_{QQ}:=QQ$&1&$\tiny \yng(2)$&2&0&$2R_v$ \\
$M_{SS}:=SS$&1&1&0&2&$2R_s$ \\
%$M_{S'S'}:=S'S'$&1&1&1&$\tiny \yng(2)$&0&0&2&$2R_c$ \\
$P_1:=SQS$&1&${\tiny \yng(1)}$&1&2&$R_v+2R_s$ \\
$P_4:=SQ^4S$&1&${\tiny \overline{\yng(1)} }$&4&2&$4R_v+2R_s$ \\
$P_5:=SQ^5S$&1&1&5&2&$5R_v+2R_s$ \\ \hline
$Z:=Y_1Y^2_2Y^2_3Y_4$&1&1&$-10$&$-4$&$2-10R_v-4R_s$ \\
$Y:=\sqrt{Y^2_1Y^2_2Y^2_3Y_4}$&1&1&$-5$&$-4$&$2-5R_v -4R_s$ \\ \hline
%$P_2:=SQ^2S$&1&1&1&1&2&2&0&$2R_v+2R_s$ \\ 
%$P'_2:=S'Q^2S'$&1&1&1&1&2&0&2&$2R_v+2R_c$ \\
%$B:=S^2S'^2$&1&1&1&1&0&2&2&$2R_s+2R_c$ \\
%$F:=S^2S'^2Q^2$&1&1&1&1&2&2&2&$2R_v+2R_s+2R_c$ \\ \hline
%$B_Q:=Q^7$&1&${\tiny \yng(1,1,1,1,1,1,1)}$&1&7&0& $7R_f$\\ 
%$B_S:=S^4$&1&1&${\tiny \yng(1,1,1,1)}$&0&$4$&$4R_S$  \\[10pt] 
%$B'_S:=S^4 Q$&1&${\tiny \yng(1)}$&${\tiny \yng(1,1,1,1)}$&1&4& $R_f+4R_S$ \\
%$P:=SQ^3S$&1&${\tiny \yng(1,1,1)}$&${\tiny \yng(2)}$&3&2& $3R_f+2R_S$ \\[9pt]
%$R:=SQ^4S$&1&${\tiny \yng(1,1,1,1)}$&${\tiny \yng(2)}$&4&2& $4R_f+2R_S$\\
%&&&&&&&& \\ \hline
%$Z:=Y_1Y_2^2Y_3Y_4$&1&1&$1$&1&$-4$&$-4$&$-4$&$2-4R_v-4R_s-4R_c$ \\  \hline
%$\det  \, M_{QQ}$&1&1&1&$2N_f$&0&$2N_fR_f$  \\
%$\det \, M_{SS}$ &1&1&1&0&$2N_S$&$2N_SR_S$  \\
%&&&&&&  \\
%&&&&&&  \\
%$\eta$&1&1&1&$2N_f$&$2N_S$&$2N_f(R_f-1)+2N_S(R_S-1) +10$  \\ \hline
%$Y_1$&1&1&1&0&$-N_S(1+\mathrm{sign}(\phi_1-\phi_3))$& $-2-N_S(R_S-1)(1+\mathrm{sign}(\phi_1-\phi_2))$\\
%$Y_2$&1&1&1&0&0&$-2$\\
%$Y_3$&1&1&1&$-2N_f$&$-N_S(1-\mathrm{sign}(\phi_1-\phi_3))$& $-2-N_S(R_S-1)(1-\mathrm{sign}(\phi_1-\phi_2))  -2N_f(R_f-1) $\\
%$Z:=Y_1Y_2^2Y_3$&1&1&1&$-2N_f$&$-2N_S$& $-8 -2N_S (R_S-1) -2N_f (R_f-1)$  \\ 
%$Y:=\sqrt{Y_1 Z}$ for $\phi_1 \ge \phi_3$ &1&1&1&$-N_f$&$-2N_S$& $-5-N_f(R_f-1) -2N_S(R_S-1)$  \\
%$Z':=Y_1Y_3$&1&1&1&$-2N_f$&$-2N_S$&$-4 -2N_S (R_S-1) -2N_f (R_f-1)$ \\
%$Z'':=Y_1Y_2Y_3$&1&1&1&$-2N_f$&$-2N_S$& $-6-2N_S (R_S-1) -2N_f (R_f-1)$  \\
%$Y_{spin}:=Y_1^2Y_2^2Y_3$&1&1&1&$-2N_f$&$-4N_S$&$-10-2N_f(R_f-1)-4N_S(R_S-1)$ \\
  \end{tabular}}
  \end{center}\label{Spin951}
\end{table}

%%%%%%%%%%%%%%%%%%%%%%%%%%%%%%%%%%%%%%%%%%%%%%%%%%%%%%%%%%
\subsection{$(N_v,N_s)=(3,2)$}
%%%%%%%%%%%%%%%%%%%%%%%%%%%%%%%%%%%%%%%%%%%%%%%%%%%%%%%%%%
The second example is the 3d $\mathcal{N}=2$ $Spin(9)$ gauge theories with three vectors and two spinors. In this case, we need not introduce the Coulomb branch coordinate $Y$ since the number of the vector matters is less than five. The Coulomb branch is one-dimensional and parametrized by $Z$. The Higgs branch operators are listed in Table \ref{Spin932}. The confining superpotential is determined from Table \ref{Spin932} as follows.
\begin{align}
W&=Z [ M_{QQ}^3 (M_{SS}^2+B)^2 +M_{QQ}^2 P_1^2  (M_{SS}^2+B)  +M_{QQ}P_2^2  (M_{SS}^2+B) \nonumber \\
& \qquad +M_{SS}P_1P_2P_3  + (P_1P_2)^2 +P_3^2 (M_{SS}^2+B)+N^2 ]
\end{align}
\begin{table}[H]\caption{3d $\mathcal{N}=2$ $Spin(9)$ theory with $(N_v,N_s)=(3,2)$} 
\begin{center}
\scalebox{1}{
  \begin{tabular}{|c||c||c|c|c|c|c| } \hline
  &$Spin(9)$&$SU(3)$&$SU(2)$&$U(1)_v$&$U(1)_s$&$U(1)_R$ \\ \hline
Q& $\mathbf{9}$&${\tiny \yng(1)}$&1&1&0& $R_v$ \\
$S$ & $\mathbf{16}$&1&${\tiny \yng(1)}$&0&1& $R_s$ \\ \hline
%$\lambda$ &$\mathbf{Adj.}$&1&1&0&0&$1$  \\ \hline
%$\eta=\Lambda_{N_v,N_S,N_c}^b$&1&1&1&1&$4$&$4$&4&$4(R_v-1)+4(R_s-1)+4(R_c-1) +12$  \\ \hline 
$M_{QQ}:=QQ$&1&$\tiny \yng(2)$&1&2&0&$2R_v$ \\
$M_{SS}:=SS$&1&1&${\tiny \yng(2)}$&0&2&$2R_s$ \\
%$M_{S'S'}:=S'S'$&1&1&1&$\tiny \yng(2)$&0&0&2&$2R_c$ \\
$P_1:=SQS$&1&${\tiny \yng(1)}$&${\tiny \yng(2)}$&1&2&$R_v+2R_s$ \\
$P_2:=SQ^2S$&1&${\tiny \overline{\yng(1)}}$&1&2&2&$2R_v+2R_s$ \\
$P_3:=SQ^3S $&1&1&1&3&2& $3R_v+2R_s$ \\
$N:=S^4 Q^3$&1&1&1&3&4&$3R_v+4R_s$ \\
$B:=S^4$&1&1&1&0&4&$4R_s$ \\ \hline
%$P_4:=SQ^4S$&1&${\tiny \overline{\yng(1)} }$&4&2&$4R_v+2R_s$ \\
%$P_5:=SQ^5S$&1&1&5&2&$5R_v+2R_s$ \\ \hline
$Z:=Y_1Y^2_2Y^2_3Y_4$&1&1&1&$-6$&$-8$&$2-6R_v-8R_s$ \\ \hline
%$Y:=\sqrt{Y^2_1Y^2_2Y^2_3Y_4}$&1&1&$-5$&$-4$&$2-5R_v -4R_s$ \\ \hline
%$P_2:=SQ^2S$&1&1&1&1&2&2&0&$2R_v+2R_s$ \\ 
%$P'_2:=S'Q^2S'$&1&1&1&1&2&0&2&$2R_v+2R_c$ \\
%$B:=S^2S'^2$&1&1&1&1&0&2&2&$2R_s+2R_c$ \\
%$F:=S^2S'^2Q^2$&1&1&1&1&2&2&2&$2R_v+2R_s+2R_c$ \\ \hline
%$B_Q:=Q^7$&1&${\tiny \yng(1,1,1,1,1,1,1)}$&1&7&0& $7R_f$\\ 
%$B_S:=S^4$&1&1&${\tiny \yng(1,1,1,1)}$&0&$4$&$4R_S$  \\[10pt] 
%$B'_S:=S^4 Q$&1&${\tiny \yng(1)}$&${\tiny \yng(1,1,1,1)}$&1&4& $R_f+4R_S$ \\
%$P:=SQ^3S$&1&${\tiny \yng(1,1,1)}$&${\tiny \yng(2)}$&3&2& $3R_f+2R_S$ \\[9pt]
%$R:=SQ^4S$&1&${\tiny \yng(1,1,1,1)}$&${\tiny \yng(2)}$&4&2& $4R_f+2R_S$\\
%&&&&&&&& \\ \hline
%$Z:=Y_1Y_2^2Y_3Y_4$&1&1&$1$&1&$-4$&$-4$&$-4$&$2-4R_v-4R_s-4R_c$ \\  \hline
%$\det  \, M_{QQ}$&1&1&1&$2N_f$&0&$2N_fR_f$  \\
%$\det \, M_{SS}$ &1&1&1&0&$2N_S$&$2N_SR_S$  \\
%&&&&&&  \\
%&&&&&&  \\
%$\eta$&1&1&1&$2N_f$&$2N_S$&$2N_f(R_f-1)+2N_S(R_S-1) +10$  \\ \hline
%$Y_1$&1&1&1&0&$-N_S(1+\mathrm{sign}(\phi_1-\phi_3))$& $-2-N_S(R_S-1)(1+\mathrm{sign}(\phi_1-\phi_2))$\\
%$Y_2$&1&1&1&0&0&$-2$\\
%$Y_3$&1&1&1&$-2N_f$&$-N_S(1-\mathrm{sign}(\phi_1-\phi_3))$& $-2-N_S(R_S-1)(1-\mathrm{sign}(\phi_1-\phi_2))  -2N_f(R_f-1) $\\
%$Z:=Y_1Y_2^2Y_3$&1&1&1&$-2N_f$&$-2N_S$& $-8 -2N_S (R_S-1) -2N_f (R_f-1)$  \\ 
%$Y:=\sqrt{Y_1 Z}$ for $\phi_1 \ge \phi_3$ &1&1&1&$-N_f$&$-2N_S$& $-5-N_f(R_f-1) -2N_S(R_S-1)$  \\
%$Z':=Y_1Y_3$&1&1&1&$-2N_f$&$-2N_S$&$-4 -2N_S (R_S-1) -2N_f (R_f-1)$ \\
%$Z'':=Y_1Y_2Y_3$&1&1&1&$-2N_f$&$-2N_S$& $-6-2N_S (R_S-1) -2N_f (R_f-1)$  \\
%$Y_{spin}:=Y_1^2Y_2^2Y_3$&1&1&1&$-2N_f$&$-4N_S$&$-10-2N_f(R_f-1)-4N_S(R_S-1)$ \\
  \end{tabular}}
  \end{center}\label{Spin932}
\end{table}

%%%%%%%%%%%%%%%%%%%%%%%%%%%%%%%%%%%%%%%%%%%%%%%%%%%%%%%%%%
\subsection{$(N_v,N_s)=(1,3)$}
%%%%%%%%%%%%%%%%%%%%%%%%%%%%%%%%%%%%%%%%%%%%%%%%%%%%%%%%%%
The final example is the 3d $\mathcal{N}=2$ $Spin(9)$ gauge theory with one vector and three spinors. In this case, the Coulomb branch is again one-dimensional and parametrized by $Z$. The Higgs branch is described by the five composite operators defined in Table \ref{Spin913}. The confining superpotential is determined as
\begin{align}
W= Z[M_{QQ}(M_{SS}^6+B^3 +M_{SS}^2B^2) +M_{SS}^4P_1^2 +(P_1B)^2+(B+M_{SS}^2)N^2  ].
\end{align}
\begin{table}[H]\caption{3d $\mathcal{N}=2$ $Spin(9)$ theory with $(N_v,N_s)=(1,3)$} 
\begin{center}
\scalebox{1}{
  \begin{tabular}{|c||c||c|c|c|c| } \hline
  &$Spin(9)$&$SU(3)$&$U(1)_v$&$U(1)_s$&$U(1)_R$ \\ \hline
Q& $\mathbf{9}$&1&1&0& $R_v$ \\
$S$ & $\mathbf{16}$&${\tiny \yng(1)}$&0&1& $R_s$ \\ \hline
%$\lambda$ &$\mathbf{Adj.}$&1&1&0&0&$1$  \\ \hline
%$\eta=\Lambda_{N_v,N_S,N_c}^b$&1&1&1&1&$4$&$4$&4&$4(R_v-1)+4(R_s-1)+4(R_c-1) +12$  \\ \hline 
$M_{QQ}:=QQ$&1&1&2&0&$2R_v$ \\
$M_{SS}:=SS$&1&${\tiny \yng(2)}$&0&2&$2R_s$ \\
%$M_{S'S'}:=S'S'$&1&1&1&$\tiny \yng(2)$&0&0&2&$2R_c$ \\
$P_1:=SQS$&1&${\tiny \yng(2)}$&1&2&$R_v+2R_s$ \\
$B:=S^4$&1&${\tiny \overline{\yng(2)}}$&0&4&$4R_s$ \\
$N:=S^4Q$&1&${\tiny \yng(1)}$&1&4&$R_v+4R_s$ \\ \hline
$Z:=Y_1Y^2_2Y^2_3Y_4$&1&1&$-2$&$-12$&$2-2R_v-12R_s$ \\ \hline
%$Y:=\sqrt{Y^2_1Y^2_2Y^2_3Y_4}$&1&1&$-5$&$-4$&$2-5R_v -4R_s$ \\ \hline
%$P_2:=SQ^2S$&1&1&1&1&2&2&0&$2R_v+2R_s$ \\ 
%$P'_2:=S'Q^2S'$&1&1&1&1&2&0&2&$2R_v+2R_c$ \\
%$B:=S^2S'^2$&1&1&1&1&0&2&2&$2R_s+2R_c$ \\
%$F:=S^2S'^2Q^2$&1&1&1&1&2&2&2&$2R_v+2R_s+2R_c$ \\ \hline
%$B_Q:=Q^7$&1&${\tiny \yng(1,1,1,1,1,1,1)}$&1&7&0& $7R_f$\\ 
%$B_S:=S^4$&1&1&${\tiny \yng(1,1,1,1)}$&0&$4$&$4R_S$  \\[10pt] 
%$B'_S:=S^4 Q$&1&${\tiny \yng(1)}$&${\tiny \yng(1,1,1,1)}$&1&4& $R_f+4R_S$ \\
%$P:=SQ^3S$&1&${\tiny \yng(1,1,1)}$&${\tiny \yng(2)}$&3&2& $3R_f+2R_S$ \\[9pt]
%$R:=SQ^4S$&1&${\tiny \yng(1,1,1,1)}$&${\tiny \yng(2)}$&4&2& $4R_f+2R_S$\\
%&&&&&&&& \\ \hline
%$Z:=Y_1Y_2^2Y_3Y_4$&1&1&$1$&1&$-4$&$-4$&$-4$&$2-4R_v-4R_s-4R_c$ \\  \hline
%$\det  \, M_{QQ}$&1&1&1&$2N_f$&0&$2N_fR_f$  \\
%$\det \, M_{SS}$ &1&1&1&0&$2N_S$&$2N_SR_S$  \\
%&&&&&&  \\
%&&&&&&  \\
%$\eta$&1&1&1&$2N_f$&$2N_S$&$2N_f(R_f-1)+2N_S(R_S-1) +10$  \\ \hline
%$Y_1$&1&1&1&0&$-N_S(1+\mathrm{sign}(\phi_1-\phi_3))$& $-2-N_S(R_S-1)(1+\mathrm{sign}(\phi_1-\phi_2))$\\
%$Y_2$&1&1&1&0&0&$-2$\\
%$Y_3$&1&1&1&$-2N_f$&$-N_S(1-\mathrm{sign}(\phi_1-\phi_3))$& $-2-N_S(R_S-1)(1-\mathrm{sign}(\phi_1-\phi_2))  -2N_f(R_f-1) $\\
%$Z:=Y_1Y_2^2Y_3$&1&1&1&$-2N_f$&$-2N_S$& $-8 -2N_S (R_S-1) -2N_f (R_f-1)$  \\ 
%$Y:=\sqrt{Y_1 Z}$ for $\phi_1 \ge \phi_3$ &1&1&1&$-N_f$&$-2N_S$& $-5-N_f(R_f-1) -2N_S(R_S-1)$  \\
%$Z':=Y_1Y_3$&1&1&1&$-2N_f$&$-2N_S$&$-4 -2N_S (R_S-1) -2N_f (R_f-1)$ \\
%$Z'':=Y_1Y_2Y_3$&1&1&1&$-2N_f$&$-2N_S$& $-6-2N_S (R_S-1) -2N_f (R_f-1)$  \\
%$Y_{spin}:=Y_1^2Y_2^2Y_3$&1&1&1&$-2N_f$&$-4N_S$&$-10-2N_f(R_f-1)-4N_S(R_S-1)$ \\
  \end{tabular}}
  \end{center}\label{Spin913}
\end{table}

%%%%%%%%%%%%%%%%%%%%%%%%%%%%%%%%%%%%%%%%%%%%%%%%%%%%%%%%%%
%%%%%%%%%%%%%%%%%%%%%%%%%%%%%%%%%%%%%%%%%%%%%%%%%%%%%%%%%%
%%%%%%%%%%%%%%%%%%%%%%%%%%%%%%%%%%%%%%%%%%%%%%%%%%%%%%%%%%
\section{$Spin(10)$ theories}
%%%%%%%%%%%%%%%%%%%%%%%%%%%%%%%%%%%%%%%%%%%%%%%%%%%%%%%%%%
%%%%%%%%%%%%%%%%%%%%%%%%%%%%%%%%%%%%%%%%%%%%%%%%%%%%%%%%%%
%%%%%%%%%%%%%%%%%%%%%%%%%%%%%%%%%%%%%%%%%%%%%%%%%%%%%%%%%%
Next, we move on to the 3d $\mathcal{N}=2$ $Spin(10)$ theory with $N_v$ vectors, $N_s$ spinors and $N_{s'}$ (complex) conjugate spinors. This case will be very special since we have to introduce a dressed Coulomb branch operator. There are three Coulomb branches where vector and spinor representations supply massless fields charged under the unbroken gauge group. The first Coulomb branch $Y$ leads to the following breaking pattern
\begin{align}
so(10)  & \rightarrow so(8)  \times u(1)  \\
\mathbf{10} & \rightarrow  \mathbf{8}_{v,0} +\mathbf{1}_{2} +\mathbf{1}_{-2} \\
\mathbf{16}   &\rightarrow   \mathbf{8}_{c,-1} +\mathbf{8}_{s,1} \\
\overline{\mathbf{16}}   &\rightarrow \mathbf{8}_{c,1} +\mathbf{8}_{s,-1}.
\end{align}
The spinor fields are all massive and integrated out. In order to make the low-energy $SO(8)$ dynamics stable, we can use $\mathbf{8}_{v,0} $ from the vector representation. Since the 3d $\mathcal{N}=2$ $SO(8)$ theory with $N_v$ vectors has a stable SUSY vacuum for $N_v \ge 6$, the $Y$-branch is available for $N_v \ge 6$. 

The second Coulomb branch $Z$ leads to the breaking
\begin{align}
so(10)  & \rightarrow so(6) \times su(2)  \times u(1)  \\
\mathbf{10} & \rightarrow  (\mathbf{6} , \mathbf{1})_{0} +(\mathbf{1} , \mathbf{2})_{\pm 1} \\
\mathbf{16}   &\rightarrow   (\mathbf{4} , \mathbf{1})_{\pm 1 }+(\overline{\mathbf{4}} , \mathbf{2})_{0} \\
\overline{\mathbf{16}}   &\rightarrow (\overline{\mathbf{4}} , \mathbf{1})_{ \pm 1} +(\mathbf{4} , \mathbf{2})_{0}.
\end{align}
In order that this branch becomes a flat direction, the vacuum of the low-energy $SO(6) \times SU(2)$ theory must have a stable SUSY vacuum. The $SU(2)$ part is made stable by $(\overline{\mathbf{4}} , \mathbf{2})_{0} \in \mathbf{16} $ or $(\mathbf{4} , \mathbf{2})_{0} \in \overline{\mathbf{16}} $. The $SO(6)$ part is made stable by both vector and spinor matters. 

The third Coulomb branch $X$ needs a special care. This operator corresponds to the gauge symmetry breaking 
\begin{align}
so(10)  & \rightarrow su(4) \times so(2)  \times u(1)  \\
\mathbf{10} & \rightarrow  \mathbf{4}_{0,-1} + \overline{\mathbf{4}}_{0,-1} +\mathbf{1}_{2,0}+\mathbf{1}_{-2,0} \\
\mathbf{16}   &\rightarrow   \mathbf{4}_{-1,-1} +\overline{\mathbf{4}}_{-1,1} +\mathbf{6}_{1,0} +\mathbf{1}_{1,2}+\mathbf{1}_{1,-2} \\
\overline{\mathbf{16}}   &\rightarrow \mathbf{4}_{1,-1} +\overline{\mathbf{4}}_{1,1} +\mathbf{6}_{-1,0} +\mathbf{1}_{-1,2}+\mathbf{1}_{-1,-2}.
\end{align}
Notice that there are two $U(1)$ factors and the Coulomb branch is related to the second $U(1)$ factor. Along this branch, the effective Chern-Simons level between $so(2)$ and $u(1)$ is introduced, which is calculated as
\begin{align}
k_{eff}^{so(2), u(1)} = -N_s +N_{s'}.
\end{align}
Therefore, the bare Coulomb branch $X$ is not gauge invariant and its $so(2)$ charge is $N_s -N_{s'}$. In order to construct a gauge invariant coordinate, we can use $\mathbf{6}_{\pm 1 ,0}$ from the spinor representation or $\mathbf{1}_{2,0}$ from the vector representation. The vacuum of the low-energy $SU(4)$ theory can be made stable only by spinor matters.

%%%%%%%%%%%%%%%%%%%%%%%%%%%%%%%%%%%%%%%%%%%%%%%%%%%%%%%%%%
\subsection{$(N_v,N_s,N_{s'})=(6,1,0)$}
%%%%%%%%%%%%%%%%%%%%%%%%%%%%%%%%%%%%%%%%%%%%%%%%%%%%%%%%%%
The first example is the 3d $\mathcal{N}=2$ $Spin(10)$ theory with six vectors and one spinor. The corresponding 4d theory was studied in \cite{Pouliot:1996zh, Kawano:1996bd}. The Higgs branch is described by three composite operators, $M_{QQ}, P_1$ and $P_5$ which are defined in Table \ref{Spin1061}. The Coulomb moduli are two-dimensional, which are parametrized by $Y$ and $Z$. The Coulomb branch operator $X$ now has an $SO(2) \simeq U(1)$ charge $2$ and cannot be made gauge invariant. The confining superpotential becomes
\begin{align}
W= Z[M_{QQ} P_5^2 +M_{QQ}^5 P_1^2] +YP_1P_5, 
\end{align}
which is consistent with the 4d result \cite{Pouliot:1996zh, Kawano:1996bd}.

\begin{table}[H]\caption{3d $\mathcal{N}=2$ $Spin(10)$ theory with $(N_v,N_s,N_{s'})=(6,1,0)$} 
\begin{center}
\scalebox{1}{
  \begin{tabular}{|c||c||c|c|c|c| } \hline
  &$Spin(10)$&$SU(6)$&$U(1)_v$&$U(1)_s$&$U(1)_R$ \\ \hline
$Q$& $\mathbf{10}$&${\tiny \yng(1)}$&1&0& $R_v$ \\
$S$ & $\mathbf{16}$&1&0&1& $R_s$ \\ \hline 
$M_{QQ}:=QQ$&1&${\tiny \yng(2)}$&2&0&$2R_v$ \\
$P_1:=SQS$&1&${\tiny \yng(1)  }$&1&2&$R_v + 2R_s$  \\
$P_{5}:=SQ^5S$ &1&${\tiny \overline{ \yng(1)}}$&5&2&$5R_v +2R_s$  \\ \hline 
$Z:=Y_1 Y_2^2 Y_3^2 Y_4 Y_5  $&1&1&$-12$&$-4$&$2-12R_v -4R_s$ \\
$Y:= \sqrt{Y_1^2 Y_2^2 Y_3^2 Y_4 Y_5 }$&1&1&$-6$&$-4$&$2 -6R_v -4R_s$  \\ \hline
  \end{tabular}}
  \end{center}\label{Spin1061}
\end{table}

%%%%%%%%%%%%%%%%%%%%%%%%%%%%%%%%%%%%%%%%%%%%%%%%%%%%%%%%%%
\subsection{$(N_v,N_s,N_{s'})=(4,2,0)$}
%%%%%%%%%%%%%%%%%%%%%%%%%%%%%%%%%%%%%%%%%%%%%%%%%%%%%%%%%%
The second example is the 3d $\mathcal{N}=2$ $Spin(10)$ theory with four vectors and two spinors.
The corresponding 4d theory was studied in \cite{Berkooz:1997bb, Kawano:2007rz}. The Coulomb branch $Y$ is not available since the low-energy $SO(8)$ theory with four vectors has no stable SUSY vacuum. The $X$-branch is also not allowed in the same manner. As a result, the Coulomb branch is one-dimensional, which is described by $Z$. Table \ref{Spin10420} shows the moduli coordinates and their quantum numbers. 
The confining superpotential becomes
\begin{align}
W=Z[B^2 \det M_{QQ} +M_{QQ}^3 P_1^2 B   +M_{QQ}^2 P_1^4+M_{QQ} P_3 P_1^3+BM_{QQ} P_3^2 +(P_1P_3)^2 +R^2],
\end{align}
which is consistent with all the symmetries in Table \ref{Spin10420} and the 4d results \cite{Berkooz:1997bb}.
\begin{table}[H]\caption{3d $\mathcal{N}=2$ $Spin(10)$ theory with $(N_v,N_s,N_{s'})=(4,2,0)$} 
\begin{center}
\scalebox{1}{
  \begin{tabular}{|c||c||c|c|c|c|c| } \hline
  &$Spin(10)$&$SU(4)$&$SU(2)$&$U(1)_v$&$U(1)_s$&$U(1)_R$ \\ \hline
Q& $\mathbf{10}$&${\tiny \yng(1)}$&1&1&0& $R_v$ \\
$S$ & $\mathbf{16}$&1&${\tiny \yng(1)}$&0&1& $R_s$ \\ \hline 
$M_{QQ}:=QQ$&1&${\tiny \yng(2)}$&1&2&0&$2R_v$ \\
$P_1:=SQS$&1&${\tiny \yng(1)  }$&${\tiny \yng(2)  }$&1&2&$R_v + 2R_s$  \\
$P_3:=SQ^3S$&1&${\tiny  \overline{\yng(1)}  }$&1&3&2&$3R_v + 2R_s$  \\
$B:=S^4$ &1&1&1&0&4&$4R_s$  \\ 
$R:=S^4 Q^4$&1&1&1&4&4&$4R_v +4R_s$ \\ \hline
$Z:=Y_1 Y_2^2 Y_3^2 Y_4 Y_5  $&1&1&1&$-8$&$-8$&$2-8R_v -8R_s$ \\ \hline
%$Y:= \sqrt{Y_1^2 Y_2^2 Y_3^2 Y_4 Y_5 }$&1&1&1&$-6$&$-4$&$2 -6R_v -4R_s$  \\ \hline
  \end{tabular}}
  \end{center}\label{Spin10420}
\end{table}

%%%%%%%%%%%%%%%%%%%%%%%%%%%%%%%%%%%%%%%%%%%%%%%%%%%%%%%%%%
\subsection{$(N_v,N_s,N_{s'})=(4,1,1)$}
%%%%%%%%%%%%%%%%%%%%%%%%%%%%%%%%%%%%%%%%%%%%%%%%%%%%%%%%%%
Let us move on to the 3d $\mathcal{N}=2$ $Spin(10)$ theory with four vectors, one spinor and one (complex) conjugate spinor. The Coulomb branch $Y$ is not allowed for the same reason as the previous example. The operator $X$ is lifted since the low-energy $SO(6) \simeq SU(4)$ theory only has two massless vectors and its vacuum is unstable. Consequently, the Coulomb branch is one-dimensional and described by $Z$.
The confining superpotential becomes
\begin{align}
W &= Z \left[ (M_{S\overline{S}}^4 +M_{S\overline{S}}^2 T_0 +T_0^2) \det M_{QQ} +M_{QQ}^3 P_1 \overline{P}_1 (T_0+M_{S\overline{S}}^2) \right. \nonumber \\
&\left. \qquad +M_{QQ}^2(P_1^2\overline{P}_1^2 +R_2^2 (T_0 +M_{S\overline{S}}^2) ) +R_2^2(R_2^2+T_4) +R_4^2(T_0+M_{S\overline{S}}^2) +(R_4 M_{S\overline{S}}+B_4 )^2  \right],
\end{align}
which is consistent with all the symmetries in Table \ref{Spin10411}.

\begin{table}[H]\caption{3d $\mathcal{N}=2$ $Spin(10)$ theory with $(N_v,N_s,N_{s'})=(4,1,1)$} 
\begin{center}
\scalebox{1}{
  \begin{tabular}{|c||c||c|c|c|c|c| } \hline
  &$Spin(10)$&$SU(4)$&$U(1)_v$&$U(1)_s$&$U(1)_{s'}$&$U(1)_R$ \\ \hline
Q& $\mathbf{10}$&${\tiny \yng(1)}$&1&0&0& $R_v$ \\
$S$ & $\mathbf{16}$&1&0&1&0& $R_s$ \\
$\overline{S}$ & $\overline{\mathbf{16}}$&1&0&0&1& $R_{s'}$ \\  \hline
$M_{QQ}:=QQ$&1&${\tiny \yng(2)}$&2&0&0&$2R_v$ \\
$M_{S\overline{S}}:= S \overline{S}$&1&1&0&1&1&$R_s +R_{s'}$ \\
$P_1:=SQS$&1&${\tiny \yng(1)  }$&1&2&0&$R_v + 2R_s$  \\
$\overline{P}_1:=\overline{S}Q\overline{S}$&1&${\tiny \yng(1)  }$&1&0&2&$R_v + 2R_{s'}$  \\
$R_2:=S Q^2\overline{S} $&1&${\tiny \yng(1,1)}$&2&1&1&$2R_v+R_s+R_{s'}$ \\
$R_4 :=S Q^4 \overline{S} $&1&1&4&1&1&$4R_v+R_s+R_{s'}$ \\
$T_0:= S^2 \overline{S}^2$&1&1&0&2&2&$2R_s+2R_{s'}$ \\
 $T_2:=S^2 \overline{S}^2 Q^4 $&1&1&4&2&2& $4R_v+2R_s+2R_{s'}$\\ \hline
$Z:=Y_1 Y_2^2 Y_3^2 Y_4 Y_5  $&1&1&$-8$&$-4$&$-4$&$2-8R_v -4R_s-4R_{s'}$ \\ \hline
%$Y:= \sqrt{Y_1^2 Y_2^2 Y_3^2 Y_4 Y_5 }$&1&1&1&$-6$&$-4$&$2 -6R_v -4R_s$  \\ \hline
  \end{tabular}}
  \end{center}\label{Spin10411}
\end{table}

%%%%%%%%%%%%%%%%%%%%%%%%%%%%%%%%%%%%%%%%%%%%%%%%%%%%%%%%%%
\subsection{$(N_v,N_s,N_{s'})=(2,3,0)$}
%%%%%%%%%%%%%%%%%%%%%%%%%%%%%%%%%%%%%%%%%%%%%%%%%%%%%%%%%
Let us consider the 3d $\mathcal{N}=2$ $Spin(10)$ theory with two vectors and three spinors. The Coulomb branch $Y$ is not allowed since the number of the vector matters is less than six. The operator $X$ is not available since the low-energy $SO(6)$ theory with three vectors has no stable SUSY vacuum. In the current case, only the $Z$-branch is available. The confined degrees of freedom are summarized in Table \ref{Spin10230}. The confining superpotential becomes
\begin{align}
W=Z [\det M_{QQ} \det B +M_{QQ} (P_1B)^2 +BR^2].
\end{align}
\begin{table}[H]\caption{3d $\mathcal{N}=2$ $Spin(10)$ theory with $(N_v,N_s,N_{s'})=(2,3,0)$} 
\begin{center}
\scalebox{1}{
  \begin{tabular}{|c||c||c|c|c|c|c| } \hline
  &$Spin(10)$&$SU(2)$&$SU(3)$&$U(1)_v$&$U(1)_s$&$U(1)_R$ \\ \hline
Q& $\mathbf{10}$&${\tiny \yng(1)}$&1&1&0& $R_v$ \\
$S$ & $\mathbf{16}$&1&${\tiny \yng(1)}$&0&1& $R_s$ \\ \hline 
$M_{QQ}:=QQ$&1&${\tiny \yng(2)}$&1&2&0&$2R_v$ \\
$P_1:=SQS$&1&${\tiny \yng(1)  }$&${\tiny \yng(2)  }$&1&2&$R_v + 2R_s$  \\
$B:=S^4$ &1&1&${\tiny  \overline{\yng(2)} }$&0&4&$4R_s$  \\ 
$R:=S^4 Q^2$&1&1&${\tiny \yng(1)}$&2&4&$2R_v +4R_s$ \\ \hline
$Z:=Y_1 Y_2^2 Y_3^2 Y_4 Y_5  $&1&1&1&$-4$&$-12$&$2-4R_v -12R_s$ \\ \hline
%$Y:= \sqrt{Y_1^2 Y_2^2 Y_3^2 Y_4 Y_5 }$&1&1&1&$-6$&$-4$&$2 -6R_v -4R_s$  \\ \hline
  \end{tabular}}
  \end{center}\label{Spin10230}
\end{table}

%%%%%%%%%%%%%%%%%%%%%%%%%%%%%%%%%%%%%%%%%%%%%%%%%%%%%%%%%%
\subsection{$(N_v,N_s,N_{s'})=(2,2,1)$}
%%%%%%%%%%%%%%%%%%%%%%%%%%%%%%%%%%%%%%%%%%%%%%%%%%%%%%%%%
The next example is the 3d $\mathcal{N}=2$ $Spin(10)$ theory with two vectors, two spinors and one conjugate spinor. As in the previous case, the Coulomb branch is described by the single operator $Z$. The Coulomb branch $Y$ is not available since $N_v$ is less than six. The Coulomb branch $X$ is not allowed since the low-energy $SO(6)$ theory with three vectors has a runaway potential.  
The moduli coordinates and their quantum numbers are summarized in Table \ref{Spin10221}. We will not explicitly write down the confining superpotential. 
\begin{table}[H]\caption{3d $\mathcal{N}=2$ $Spin(10)$ theory with $(N_v,N_s,N_{s'})=(2,2,1)$} 
\begin{center}
\scalebox{0.9}{
  \begin{tabular}{|c||c||c|c|c|c|c|c| } \hline
  &$Spin(10)$&$SU(2)$&$SU(2)$&$U(1)_v$&$U(1)_s$&$U(1)_{s'}$&$U(1)_R$ \\ \hline
Q& $\mathbf{10}$&${\tiny \yng(1)}$&1&1&0&0& $R_v$ \\
$S$ & $\mathbf{16}$&1&${\tiny \yng(1)}$&0&1&0& $R_s$ \\ 
$\overline{S}$ & $\overline{\mathbf{16}}$&1&1&0&0&1& $R_s$ \\  \hline
$M_{QQ}:=QQ$&1&${\tiny \yng(2)}$&1&2&0&0&$2R_v$ \\
$M_{S\overline{S}}:=S\overline{S}$&1&1&${\tiny \yng(1)}$&0&1&1&$R_s+R_{s'}$ \\
$M_{2,S\overline{S}}:=SQ^2\overline{S}$&1&1&${\tiny \yng(1)}$&2&1&1&$2R_v+R_s+R_{s'}$ \\
$P_1:=SQS$&1&${\tiny \yng(1)  }$&${\tiny \yng(2)  }$&1&2&0&$R_v + 2R_s$  \\
$\overline{P}_1:=\overline{S}Q\overline{S}$&1&${\tiny \yng(1)  }$&1&1&0&2&$R_v + 2R_{s'}$  \\
$B:=S^4$&1&1&1&0&4&0&$4R_s$ \\
$F:=S^2 \overline{S}^2$ &1&1&${\tiny \yng(2) }$&0&2&2&$2R_s+2R_{s'}$  \\ 
$R_1:=S^3 \overline{S} Q$&1&${\tiny \yng(1)}$&${\tiny \yng(1)}$&1&3&2&$R_v +3R_s+R_{s'}$ \\ 
$R_2:=S^2 \overline{S}^2 Q^2$&1&1&1&2&2&2&$2R_v +2R_s+2R_{s'}$ \\  \hline
$Z:=Y_1 Y_2^2 Y_3^2 Y_4 Y_5  $&1&1&1&$-4$&$-8$&$-4$&$2-4R_v -8R_s -4 R_{s'}$ \\ \hline
%$Y:= \sqrt{Y_1^2 Y_2^2 Y_3^2 Y_4 Y_5 }$&1&1&1&$-6$&$-4$&$2 -6R_v -4R_s$  \\ \hline
  \end{tabular}}
  \end{center}\label{Spin10221}
\end{table}

%%%%%%%%%%%%%%%%%%%%%%%%%%%%%%%%%%%%%%%%%%%%%%%%%%%%%%%%%%
\subsection{$(N_v,N_s,N_{s'})=(0,4,0)$}
%%%%%%%%%%%%%%%%%%%%%%%%%%%%%%%%%%%%%%%%%%%%%%%%%%%%%%%%%
Next, we move on to the theories with spinor matters and without a vector. The first example of the s-confinement is the 3d $\mathcal{N}=2$ $Spin(10)$ gauge theory with four spinors. Since the theory does not include the vector matters, the Coulomb branch $Y$ is not available. The direction $Z$ can be made stable by the component $(\overline{\mathbf{4}} , \mathbf{2})_{0} \in \mathbf{16}$. The Coulomb branch $X$ is now charged under the $so(2)$ subgroup and cannot be made gauge invariant since there is no complex conjugate spinor ($\overline{\mathbf{16}}$) in the theory. As a result, the Coulomb branch is one-dimensional and described by $Z$. The confining superpotential becomes
\begin{align}
W= Z B^4,
\end{align}
which is consistent with all the symmetries in Table \ref{Spin10040}.
\begin{table}[H]\caption{3d $\mathcal{N}=2$ $Spin(10)$ theory with $(N_v,N_s,N_{s'})=(0,4,0)$} 
\begin{center}
\scalebox{1}{
  \begin{tabular}{|c||c||c|c|c| } \hline
  &$Spin(10)$&$SU(4)$&$U(1)_s$&$U(1)_R$ \\ \hline
%Q& $\mathbf{10}$&${\tiny \yng(1)}$&1&0& $R_v$ \\
$S$ & $\mathbf{16}$&${\tiny \yng(1)}$&1& $R_s$ \\  \hline
%$\overline{S}$&$\overline{\mathbf{16}}$&1&0&1&$R_{s'}$\\ \hline
$B:=S^4$&1&${\tiny \yng(2,2)}$&4&$4R_s$ \\ \hline 
$Z:=Y_1 Y_2^2 Y_3^2 Y_4 Y_5 $&1&1&$-16$&$2-16R_{s}$ \\ 
\hline
  \end{tabular}}
  \end{center}\label{Spin10040}
\end{table}

%%%%%%%%%%%%%%%%%%%%%%%%%%%%%%%%%%%%%%%%%%%%%%%%%%%%%%%%%%
\subsection{$(N_v,N_s,N_{s'})=(0,3,1)$}
%%%%%%%%%%%%%%%%%%%%%%%%%%%%%%%%%%%%%%%%%%%%%%%%%%%%%%%%%
Let us consider the 3d $\mathcal{N}=2$ $Spin(10)$ theory with three spinors and a single (complex) conjugate spinor. The Coulomb branch $Z$ is available since the low-energy $SU(4)$ theory with two fundamentals and six anti-fundamentals has a stable SUSY vacuum \cite{Nii:2018bgf}. Similarly, the Coulomb branch $X$ is allowed although it is not gauge invariant. Therefore, we need to introduce the dressed operator 
\begin{align}
X^{dressed} := X {\overline{S}}^2 .
\end{align}
The moduli coordinates and their quantum numbers are summarized in Table \ref{Spin10031}. The confining superpotential becomes
\begin{align}
W= Z\left[ B^2 (F_2+M_{S \overline{S}}^2)^2 +C^2   (F_2+M_{S \overline{S}}^2)\right] +X^{dressed}  BC.
\end{align}
\begin{table}[H]\caption{3d $\mathcal{N}=2$ $Spin(10)$ theory with $(N_v,N_s,N_{s'})=(0,3,1)$} 
\begin{center}
\scalebox{1}{
  \begin{tabular}{|c||c||c|c|c|c| } \hline
  &$Spin(10)$&$SU(3)$&$U(1)_s$&$U(1)_{s'}$&$U(1)_R$ \\ \hline
%Q& $\mathbf{10}$&${\tiny \yng(1)}$&1&0& $R_v$ \\
$S$ & $\mathbf{16}$&${\tiny \yng(1)}$&1&0& $R_s$ \\ 
$\overline{S}$&$\overline{\mathbf{16}}$&1&0&1&$R_{s'}$\\ \hline
$M_{S\overline{S}}:= S \overline{S}$&1&${\tiny \yng(1)}$&1&1&$R_s+R_{s'}$ \\
$F_2:=S^2 \overline{S}^2$&1&${\tiny \yng(2)  }$&2&2&$2R_s + 2R_{s'}$  \\
$B:=S^4$ &1&${\tiny \overline{ \yng(2)}}$&4&0&$4R_s$  \\ 
$C:=S^5 \overline{S} $&1&${\tiny \yng(2)  }$&5&1&$5R_s + R_{s'}$ \\ \hline 
$Z:=Y_1 Y_2^2 Y_3^2 Y_4 Y_5  $&1&1&$-12$&$-4$&$2-12R_s -4R_{s'}$ \\
$X^{dressed}:= {\overline{S}}^2 \sqrt{Y_1 Y_2^2 Y_3^3Y_4^2Y_5^2}$&1&1&$-9$&$-1$&$2 -9R_s -R_{s'}$  \\ \hline
  \end{tabular}}
  \end{center}\label{Spin10031}
\end{table}

%%%%%%%%%%%%%%%%%%%%%%%%%%%%%%%%%%%%%%%%%%%%%%%%%%%%%%%%%%
\subsection{$(N_v,N_s,N_{s'})=(0,2,2)$}
%%%%%%%%%%%%%%%%%%%%%%%%%%%%%%%%%%%%%%%%%%%%%%%%%%%%%%%%%
The final example is the 3d $\mathcal{N}=2$ $Spin(10)$ theory with two spinors and two (complex) conjugate spinors. The theory is ``vector-like'' in the sense that there are equal number of spinors and conjugate spinors. Since the theory is now ``vector-like'', the bare Coulomb branch operator $X$ is gauge invariant and does not need ``dressing''. The low-energy $SU(4) \simeq SO(6)$ theory along $\braket{X} \neq 0$ contains four vector matters and hence its low-energy vacuum is stable and supersymmetric. The Coulomb branch $Z$ is also allowed since the low-energy $SU(4)$ theory with four fundamental flavors has a stable SUSY vacuum. Table \ref{Spin10022} summarizes the quantum numbers of the moduli coordinates. 

\begin{table}[H]\caption{3d $\mathcal{N}=2$ $Spin(10)$ theory with $(N_v,N_s,N_{s'})=(0,2,2)$} 
\begin{center}
\scalebox{1}{
  \begin{tabular}{|c||c||c|c|c|c|c| } \hline
  &$Spin(10)$&$SU(2)_s$&$SU(2)_{s'}$&$U(1)_s$&$U(1)_{s'}$&$U(1)_R$ \\ \hline
%Q& $\mathbf{10}$&${\tiny \yng(1)}$&1&0&0& $R_v$ \\
$S$ & $\mathbf{16}$&${\tiny \yng(1)}$&1&1&0& $R_s$ \\
$\overline{S}$ & $\overline{\mathbf{16}}$&1&${\tiny \yng(1)}$&0&1& $R_{s'}$ \\  \hline
%$M_{QQ}:=QQ$&1&${\tiny \yng(2)}$&2&0&0&$2R_v$ \\
$M_{S\overline{S}}:= S \overline{S}$&1&${\tiny \yng(1)}$&${\tiny \yng(1)}$&1&1&$R_s +R_{s'}$ \\
$B:=S^4$&1&1&1&4&0&$4R_s$ \\
$\overline{B}:=\overline{S}^4$&1&1&1&0&4&$4R_{s'}$ \\
$F_2 :=S^2 \overline{S}^2$&1&${\tiny \yng(2)}$&${\tiny \yng(2)}$&2&2&$2R_s+2R_{s'}$ \\
$F_3:=S^3 \overline{S}^3$&1&${\tiny \yng(1)}$&${\tiny \yng(1)}$&3&3&$3R_s +3R_{s'}$ \\
$C_{6,2}=S^6 \overline{S}^2$&1&1&1&6&2&$6R_s+2R_{s'}$ \\
$C_{2,6}:=S^2 \overline{S}^6$&1&1&1&2&6&$2R_s+6R_{s'}$ \\ \hline
$Z:=Y_1 Y_2^2 Y_3^2 Y_4 Y_5  $&1&1&$1$&$-8$&$-8$&$2-8R_s-8R_{s'}$ \\ 
$X:=\sqrt{Y_1 Y_2^2 Y_3^3Y_4^2Y_5^2 }$&1&1&1&$-6$&$-6$&$2-6R_s -6R_{s'}$ \\ \hline
%$Y:= \sqrt{Y_1^2 Y_2^2 Y_3^2 Y_4 Y_5 }$&1&1&1&$-6$&$-4$&$2 -6R_v -4R_s$  \\ \hline
  \end{tabular}}
  \end{center}\label{Spin10022}
\end{table}

%%%%%%%%%%%%%%%%%%%%%%%%%%%%%%%%%%%%%%%%%%%%%%%%%%%%%%%%%%
%%%%%%%%%%%%%%%%%%%%%%%%%%%%%%%%%%%%%%%%%%%%%%%%%%%%%%%%%%
%%%%%%%%%%%%%%%%%%%%%%%%%%%%%%%%%%%%%%%%%%%%%%%%%%%%%%%%%%
\section{$Spin(11)$ theories}
%%%%%%%%%%%%%%%%%%%%%%%%%%%%%%%%%%%%%%%%%%%%%%%%%%%%%%%%%%
%%%%%%%%%%%%%%%%%%%%%%%%%%%%%%%%%%%%%%%%%%%%%%%%%%%%%%%%%%
%%%%%%%%%%%%%%%%%%%%%%%%%%%%%%%%%%%%%%%%%%%%%%%%%%%%%%%%%%
Here, we consider the 3d $\mathcal{N}=2$ $Spin(11)$ theory with $N_v$ vectors and $N_s$ spinors. The correponding 4d theory was studied in \cite{Cho:1997sa}. As will be explained in the following subsections, the s-confinement phases appear in $(N_v, N_s) =(5,1)$ and $(N_v, N_s) =(1,2)$. There are three Coulomb branches whose branching rules include the fields neutral under the unbroken $U(1)$ subgroup but charged under the non-abelian subgroups. The first Coulomb branch $Y$ corresponds to the breaking
\begin{align}
so(11) & \rightarrow so(9) \times u(1) \\
\mathbf{11} & \rightarrow \mathbf{9}_0 +\mathbf{1}_2 +\mathbf{1}_{-2} \\
\mathbf{32} &  \rightarrow  \mathbf{16}_1 +\mathbf{16}_{-1},
\end{align}
where all the components of the spinor representation are massive and those masses are proportional to the $U(1)$ charges. The vector field reduces to the massless $\mathbf{9}$ representation. When the $Spin(11)$ theory has more than six vectors, the vacuum of the low-energy $SO(9)$ theory can be stable and supersymmetric due to the sufficient number of $\mathbf{9}$ vectors. In the s-confining examples which will be discussed in the following subsections, the theory contains $N_v \le 5$ vectors. Therefore, this branch does not appear in what follows. See \cite{Aharony:2011ci}, where  the 3d $\mathcal{N}=2$ $SO(11)$ theory with $N_v$ vectors is studied and this operator is introduced.

When the second Coulomb branch $Z$ obtains an expectation value, the gauge group is broken as
\begin{align}
so(11) & \rightarrow so(7) \times su(2) \times u(1) \\
\mathbf{11} & \rightarrow (\mathbf{7},\mathbf{1})_{0} + (\mathbf{1},\mathbf{2})_{ \pm 1} \\
\mathbf{32} &  \rightarrow  (\mathbf{8},\mathbf{2})_{0} +(\mathbf{8},\mathbf{1})_{\pm 1}.
\end{align}
Along this direction, the $Spin(11)$ theory must have at least one spinor so that the vacuum of the low-energy $SU(2)$ theory has a stable supersymmetric vacuum. Otherwise, this direction is quantum-mechanically lifted and excluded from the chiral ring. In order to make the vacuum of the low-energy $SO(7)$ theory stable, we have to take $(N_v ,N_s)$ above the s-confinement bound of the $Spin(7)$ theory, which was studied in \cite{Nii:2018tnd}.

The third Coulomb branch $X$ corresponds to the breaking 
\begin{align}
so(11) & \rightarrow so(3) \times su(4) \times u(1) \\
\mathbf{11} & \rightarrow (\mathbf{3} , \mathbf{1})_{0}+(\mathbf{1} , \mathbf{4})_{1}+(\mathbf{1} , \overline{\mathbf{4}})_{-1} \\
\mathbf{32} &  \rightarrow  (\mathbf{2} , \mathbf{6})_{0} +(\mathbf{2} , \mathbf{1})_{-2}+(\mathbf{2} , \mathbf{1})_{-2} +(\mathbf{2} , \mathbf{4})_{-1}+(\mathbf{2} , \overline{\mathbf{4}})_{1}.
\end{align}
When there are two spinor matters, the low-energy $SU(4)$ dynamics is stable by the two massless components $(\mathbf{2} , \mathbf{6})_{0}$. The $SO(3)$ vacuum can be made stable by $(\mathbf{3} , \mathbf{1})_{0}$ or $(\mathbf{2} , \mathbf{6})_{0}$. Therefore, the $Spin(11)$ theory with more than one spinor includes this branch.

%%%%%%%%%%%%%%%%%%%%%%%%%%%%%%%%%%%%%%%%%%%%%%%%%%%%%%%%%%
\subsection{$(N_v,N_s)=(5,1)$}
%%%%%%%%%%%%%%%%%%%%%%%%%%%%%%%%%%%%%%%%%%%%%%%%%%%%%%%%%%
The first s-confining example is the 3d $\mathcal{N}=2$ $Spin(11)$ gauge theory with five vectors and one spinor. The corresponding 4d theory was studied in \cite{Cho:1997sa}. Since the number of the vector matters is less than seven, the Coulomb branch $Y$ is not available. The $X$-branch is also not required since a single spinor $(\mathbf{2} , \mathbf{6})_{0} \in \mathbf{16}$ cannot make the low-energy $SU(4) \simeq SO(6)$ vacuum stable. As a result, there is a one-dimensional Coulomb branch parametrized by $Z$.

The low-energy dynamics is dual to a non-gauge theory with the Higgs branch fields $M_{QQ}, B, P_1, P_2, R$ and the Coulomb branch field $Z$. Table \ref{Spin1151} shows the quantum numbers of these moduli fields. The confining superpotential takes
\begin{align}
W&=Z \left[ B^2 \det M_{QQ} +B M_{QQ}^4 P_1^2 +B M_{QQ}^3 P_2^2 \right. \nonumber \\
&  \qquad  \qquad \left.+M_{QQ}^2 P_1^2 P_2^2 +M_{QQ}P_2^4 +P_1P_2^2 P_5 +BP_5^2 +R^2   \right].
\end{align}
When we put the 4d theory on $\mathbb{S}^1 \times \mathbb{R}^3$, an additional non-perturbative superpotential $\Delta W= \eta Z$ is added to the above superpotential. By integrating out the Coulomb branch operator, we can reproduce the quantum-mechanically deformed moduli space in the 4d $\mathcal{N}=1$ $Spin(11)$ theory with five vectors and one spinor \cite{Cho:1997sa}.

\begin{table}[H]\caption{3d $\mathcal{N}=2$ $Spin(11)$ theory with $(N_v,N_s)=(5,1)$} 
\begin{center}
\scalebox{1}{
  \begin{tabular}{|c||c||c|c|c|c| } \hline
  &$Spin(11)$&$SU(5)$&$U(1)_v$&$U(1)_s$&$U(1)_R$ \\ \hline
Q& $\mathbf{11}$&${\tiny \yng(1)}$&1&0& $R_v$ \\
$S$ & $\mathbf{32}$&1&0&1& $R_s$ \\ \hline 
$M_{QQ}:=QQ$&1&${\tiny \yng(2)}$&2&0&$2R_v$ \\
$B:=S^4$ &1&1&0&4&$4R_s$ \\
$P_1:=SQS$&1&${\tiny \yng(1) }$&1&2&$R_v + 2R_s$  \\
$P_2:=SQ^2S$&1&${\tiny \yng(1,1)  }$&2&2&$2R_v + 2R_s$  \\
$P_5:=SQ^5S$&1&1&5&2&$5R_v + 2R_s$  \\
$R:=S^4Q^5$&1&1&5&4&$5R_v +4R_s$ \\ \hline 
$Z:=Y_1 Y_2^2 Y_3^2 Y_4^2 Y_5  $&1&1&$-10$&$-8$&$2-10R_v -8R_s$ \\ \hline
%$X:= \sqrt{Y_1 Y_2^2 Y_3^3 Y_4^4 Y_5^4 Y_6^2 }$&1&1&$-6$&$-12$&$2 -6R_v -12R_s$  \\ \hline
  \end{tabular}}
  \end{center}\label{Spin1151}
\end{table}

%%%%%%%%%%%%%%%%%%%%%%%%%%%%%%%%%%%%%%%%%%%%%%%%%%%%%%%%%%
\subsection{$(N_v,N_s)=(1,2)$}
%%%%%%%%%%%%%%%%%%%%%%%%%%%%%%%%%%%%%%%%%%%%%%%%%%%%%%%%%%
The second example is the 3d $\mathcal{N}=2$ $Spin(11)$ gauge theory with one vector and two spinors. The $Y$-branch is not available since there is only a single vector which is insufficient for the stable $SO(9)$ vacuum. The $Z$-branch is required since the $SO(7)$ vacuum is made stable by $(\mathbf{7},\mathbf{1})_{0}$ and two $(\mathbf{8},\mathbf{2})_{0}$. 
In addition to $Z$, the $X$-branch can be now turned on since the $SO(6) \simeq SU(4)$ theory with four vectors $\mathbf{6}$ can have a stable SUSY vacuum. The Coulomb branch is two-dimensional and the Higgs branch is described by the fields listed in Table \ref{Spin1112}. We will not explicitly write down the confining potential but one can construct it from Table \ref{Spin1112}.

\begin{table}[H]\caption{3d $\mathcal{N}=2$ $Spin(11)$ theory with $(N_v,N_s)=(1,2)$} 
\begin{center}
\scalebox{1}{
  \begin{tabular}{|c||c||c|c|c|c| } \hline
  &$Spin(11)$&$SU(2)$&$U(1)_v$&$U(1)_s$&$U(1)_R$ \\ \hline
Q& $\mathbf{11}$&1&1&0& $R_v$ \\
$S$ & $\mathbf{32}$&${\tiny \yng(1)}$&0&1& $R_s$ \\ \hline 
$M_{QQ}:=QQ$&1&1&2&0&$2R_v$ \\
$M_{SS}:=SS$&1&1&0&2&$2R_s$ \\
$B:=S^4$ &1&${\tiny \yng(4)}$&0&4&$4R_s$ \\
$B':=S^4$ &1&1&0&4&$4R_s$ \\
$P_1:=SQS$&1&${\tiny \yng(2) }$&1&2&$R_v + 2R_s$  \\
$F_1:=S^4 Q$&1&${\tiny \yng(2) }$&1&4&$R_v+4R_s$ \\
$F'_1:=S^4 Q$&1&1&1&4&$R_v+4R_s$ \\
$F_2:=S^4 Q^2$&1&1&2&4&$2R_v+4R_s$ \\
$T_0:=S^6 $&1&1&0&6&$6R_s$ \\
$T_1:=S^6 Q$&1&${\tiny \yng(2) }$&1&6&$R_v+6R_s$ \\
$U_0 :=S^8$&1&1&0&8&$8R_s$ \\
$U_1 :=S^8 Q$&1&1&1&8&$R_v+8R_s$ \\ \hline
$Z:=Y_1 Y_2^2 Y_3^2 Y_4^2 Y_5  $&1&1&$-2$&$-16$&$2-2R_v -16R_s$ \\ 
$X:= \sqrt{Y_1 Y_2^2 Y_3^3 Y_4^4 Y_5^2 }$&1&1&$-2$&$-12$&$2 -2R_v -12R_s$  \\ \hline
  \end{tabular}}
  \end{center}\label{Spin1112}
\end{table}

%%%%%%%%%%%%%%%%%%%%%%%%%%%%%%%%%%%%%%%%%%%%%%%%%%%%%%%%%%
%%%%%%%%%%%%%%%%%%%%%%%%%%%%%%%%%%%%%%%%%%%%%%%%%%%%%%%%%%
%%%%%%%%%%%%%%%%%%%%%%%%%%%%%%%%%%%%%%%%%%%%%%%%%%%%%%%%%%
\section{$Spin(12)$ theories}
%%%%%%%%%%%%%%%%%%%%%%%%%%%%%%%%%%%%%%%%%%%%%%%%%%%%%%%%%%
%%%%%%%%%%%%%%%%%%%%%%%%%%%%%%%%%%%%%%%%%%%%%%%%%%%%%%%%%%
%%%%%%%%%%%%%%%%%%%%%%%%%%%%%%%%%%%%%%%%%%%%%%%%%%%%%%%%%%
Let us move on to the 3d $\mathcal{N}=2$ $Spin(12)$ theory with $N_v$ vectors, $N_s$ (Weyl) spinors and $N_{s'}$ conjugate (another Weyl) spinors. The correponding 4d theory was studied, for instance, in \cite{Maru:1998hp}. We will find three s-confinement examples for $(N_v,N_s,N_{s'}) =(6,1,0),(2,2,0),(2,1,0)$. In this case, various directions of the classical Coulomb branches can be stable and survive quantum corrections since we have two inequivalent spinors and the branching rules of these spinors are different. We start with the $Y$ direction whose expectation value leads to the breaking 
\begin{align}
so(12) & \rightarrow so(10) \times u(1) \\
\mathbf{12} & \rightarrow \mathbf{10}_0 +\mathbf{1}_2 +\mathbf{1}_{-2}  \\
\mathbf{32} & \rightarrow \mathbf{16}_{-1} +\overline{\mathbf{16}}_1 \\
\mathbf{32'} & \rightarrow \mathbf{16}_{1} +\overline{\mathbf{16}}_{-1}.
\end{align}
Since the spinor matters are massive along this direction, the $Spin(12)$ theory with only spinors cannot have this branch as a flat direction. In order to make the vacuum of the low-energy $SO(10)$ theory stable, the theory must have $N_v \ge 8$ vector matters. In this section, we will consider the cases with $N_v \le 6$ and then this operator does not appear in the following discussion. 

The second Coulomb branch $Z$ corresponds to the breaking
\begin{align}
so(12) & \rightarrow so(8) \times su(2) \times u(1) \\
\mathbf{12} & \rightarrow  (\mathbf{8}_v ,\mathbf{1})_0 +(\mathbf{1},\mathbf{2})_{\pm 1}  \\
\mathbf{32} & \rightarrow  (\mathbf{8}_s, \mathbf{1})_{\pm 1} +(\mathbf{8}_c ,\mathbf{2})_0   \\
\mathbf{32'} & \rightarrow (\mathbf{8}_c, \mathbf{1})_{\pm 1} +(\mathbf{8}_s ,\mathbf{2})_0 
\end{align}
The $SU(2)$ dynamics can be made stable and supersymmetric by the components $(\mathbf{8}_c ,\mathbf{2})_0$ or $(\mathbf{8}_s ,\mathbf{2})_0$. 
The $SO(8)$ vacuum can be made stable by  $(\mathbf{8}_v ,\mathbf{1})_0$, $(\mathbf{8}_c ,\mathbf{2})_0 $ or $(\mathbf{8}_s ,\mathbf{2})_0$. In all the s-confinement examples which we discuss in the following subsections, there are enough $\mathbf{8}$ dimensional representations so that this branch becomes a quantum moduli operator.

The third Coulomb branch $X$ corresponds to the following breaking
\begin{align}
so(12) & \rightarrow so(4) \times su(4) \times u(1)  \\
\mathbf{12} & \rightarrow (\mathbf{4},\mathbf{1})_0 +  (\mathbf{1},\mathbf{4})_1+(\mathbf{1}, \overline{\mathbf{4}})_{-1}   \\
\mathbf{32} & \rightarrow  (\mathbf{2},\mathbf{6})_0  +(\mathbf{2},\mathbf{1})_2+(\mathbf{2},\mathbf{1})_{-2}  +(\mathbf{2}^*,\mathbf{4})_{-1} +(\mathbf{2}^*, \overline{\mathbf{4}})_1  \\
\mathbf{32'} & \rightarrow   (\mathbf{2}^*,\mathbf{6})_0 +(\mathbf{2}^*,\mathbf{1})_2 +(\mathbf{2}^*,\mathbf{1})_{-2} +(\mathbf{2},\mathbf{4})_{-1} +(\mathbf{2}, \overline{\mathbf{4}})_1.
\end{align}
The vacuum of the $SO(4)$ dynamics can be made stable by the first components of the above branching rules, which are neutral under the $U(1)$ subgroup and hence massless. In order to have a stable SUSY vacuum of the $SU(4) \sim SO(6)$ part, we need at least four $\mathbf{6}$ representations. Therefore, the $Spin(12)$ theories with two spinors or more will contain the $X$ operator in their spectrum of the chiral ring.

The final Coulomb branch $V$ corresponds to the breaking
\begin{align}
so(12) & \rightarrow su(6) \times u(1)  \\
\mathbf{12} & \rightarrow  \mathbf{6}_1 + \overline{\mathbf{6}}_{-1}  \\
\mathbf{32} & \rightarrow \mathbf{20}_0+  \mathbf{6}_{-2} + \overline{\mathbf{6}}_{2} \\
\mathbf{32'} & \rightarrow    \mathbf{15}_{-1} +\overline{\mathbf{15}}_{1}+\mathbf{1}_3 +\mathbf{1}_{-3}.
\end{align}
Almost all the components are massive while the spinor field leads to a massless third-order antisymmetric tensor of the unbroken $SU(6)$, which can make the $SU(6)$ vacuum stable. As studied in \cite{Nii:2018erm}, the $SU(6)$ theory with a single three-index matter cannot have a stable vacuum, which will lead to a runaway potential. Therefore, the $Spin(12)$ theory with more than one spinor can have this direction as a quantum flat direction.

%%%%%%%%%%%%%%%%%%%%%%%%%%%%%%%%%%%%%%%%%%%%%%%%%%%%%%%%%%
\subsection{$(N_v,N_s,N_{s'})=(6,1,0)$}
%%%%%%%%%%%%%%%%%%%%%%%%%%%%%%%%%%%%%%%%%%%%%%%%%%%%%%%%%
The first s-confinement example is the 3d $\mathcal{N}=2$ $Spin(12)$ theory with six vectors and one spinor. The $Y$ operator is not allowed since the low-energy $SO(10)$ theory along this direction contains only six $\mathbf{10}$ representations, which generates a runaway potential and this vacuum is unstable. Along the $Z$ direction, the low-energy $SO(8)$ dynamics is made stable by $(\mathbf{8}_v ,\mathbf{1})_0 \in \mathbf{12}$ and the $SU(2)$ part is also made stable by $(\mathbf{8}_c ,\mathbf{2})_0 \in \mathbf{32}$.  
The $X$ direction is unstable since the $SU(4) \sim SO(6)$ theory only contains two $\mathbf{6}$ representations, which is insufficient for a stable supersymmetric vacuum. The $V$ direction is also excluded due to the similar reason. The confinement phase is described by the five Higgs branch operators defined in Table \ref{Spin1261} and a single Coulomb branch $Z$. The superpotential becomes 
\begin{align}
W= Z \left[ B^2 \det M_{QQ} + P_6 \mathrm{Pf} \, P_2 +M_{QQ}^4 P_2^2 B+M_{QQ}^2 P_2^4 +BP_6^2 +F^2 \right]. \label{Spin1261W}
\end{align}
The quantum numbers of the moduli operators are summarized in Table \ref{Spin1261}. The corresponding 4d theory was studied in \cite{Maru:1998hp} and \eqref{Spin1261W} is consistent with the 4d result where we have a quantum-deformed constraint.
 
\begin{table}[H]\caption{3d $\mathcal{N}=2$ $Spin(12)$ theory with $(N_v,N_s,N_{s'})=(6,1,0)$} 
\begin{center}
\scalebox{1}{
  \begin{tabular}{|c||c||c|c|c|c| } \hline
  &$Spin(12)$&$SU(6)$&$U(1)_v$&$U(1)_s$&$U(1)_R$ \\ \hline
Q& $\mathbf{12}$&${\tiny \yng(1)}$&1&0& $R_v$ \\
$S$ & $\mathbf{32}$&1&0&1& $R_s$ \\ \hline 
$M_{QQ}:=QQ$&1&${\tiny \yng(2)}$&2&0&$2R_v$ \\
$P_2:=SQ^2S$&1&${\tiny \yng(1,1) }$&2&2&$2R_v + 2R_s$  \\
$P_6:=SQ^6S$&1&1&6&2&$6R_v + 2R_s$  \\
$B:=S^4$&1&1&0&4&$4R_s$ \\
$F:=S^4Q^6$&1&1&6&4&$6R_v +4R_s$ \\ \hline 
$Z:=Y_1 Y_2^2 Y_3^2 Y_4^2 Y_5 Y_6 $&1&1&$-12$&$-8$&$2-12R_v -8R_s$ \\ \hline
%$X:= \sqrt{Y_1 Y_2^2 Y_3^3 Y_4^4 Y_5^4 Y_6^2 }$&1&1&$-6$&$-12$&$2 -6R_v -12R_s$  \\ \hline
  \end{tabular}}
  \end{center}\label{Spin1261}
\end{table}

%%%%%%%%%%%%%%%%%%%%%%%%%%%%%%%%%%%%%%%%%%%%%%%%%%%%%%%%%%
\subsection{$(N_v,N_s,N_{s'})=(2,2,0)$}
%%%%%%%%%%%%%%%%%%%%%%%%%%%%%%%%%%%%%%%%%%%%%%%%%%%%%%%%%
The next s-confinement example is the 3d $\mathcal{N}=2$ $Spin(12)$ theory with two vectors and two spinors. In this case, the $Y$ branch is not allowed as in the previous case. The Coulomb branch $Z$ becomes stable since the low-energy $Spin(8)$ theory has two vectors and four spinors and it leads to a stable vacuum. Along the $X$-branch, the low-energy $SU(4) \simeq SO(6)$ theory contains four vectors and its vacuum is stable and supersymmetric. The $V$ direction is also allowed since the low-energy $SU(6)$ theory contains two third-order antisymmetric matters and can become stable. As a result, the Coulomb branch is now three-dimensional and described by $Z, X$ and $V$. We will not explicitly show the confining superpotential. Table \ref{Spin12220} shows the moduli fields and their quantum numbers. 
One can write down the superpotential from Table \ref{Spin12220}.

\begin{table}[H]\caption{3d $\mathcal{N}=2$ $Spin(12)$ theory with $(N_v,N_s,N_{s'})=(2,2,0)$} 
\begin{center}
\scalebox{0.99}{
  \begin{tabular}{|c||c||c|c|c|c|c| } \hline
  &$Spin(12)$&$SU(2)$&$SU(2)$&$U(1)_v$&$U(1)_s$&$U(1)_R$ \\ \hline
Q& $\mathbf{12}$&${\tiny \yng(1)}$&1&1&0& $R_v$ \\
$S$ & $\mathbf{32}$&1&${\tiny \yng(1)}$&0&1& $R_s$ \\ \hline
%$S'$ & $\mathbf{32'}$&1&0&0&1& $R_{s'}$ \\
$M_{QQ}:=QQ$&1&${\tiny \yng(2)}$&1&2&0&$2R_v$ \\
$M_{SS}:=SS$&1&1&1&0&2&$2R_v$ \\
$P_2:=SQ^2S$&1&1&${\tiny \yng(2) }$&2&2&$2R_v + 2R_s$  \\
%$P_6:=SQ^6S$&1&1&6&2&$6R_v + 2R_s$  \\
$B_0:=S^4$&1&1&${\tiny \yng(4) }$&0&4&$4R_s$ \\
$B_2:=S^4Q^2$&1&1&${\tiny \yng(2) }$&2&4&$2R_v+4R_s$ \\
$B_2':=S^4Q^2$&1&${\tiny \yng(2) }$&1&2&4&$2R_v+4R_s$ \\
$F_0:=S^6$&1&1&1&0&6&$6R_s$ \\ 
$F_2:=S^6Q^2$&1&1&${\tiny \yng(2) }$&2&6&$2R_v+6R_s$ \\ 
$T_2:=S^8 Q^2$&1&${\tiny \yng(2) }$&1&2&8&$2R_v+8R_s$ \\  \hline
$Z:=Y_1 Y_2^2 Y_3^2 Y_4^2 Y_5 Y_6 $&1&1&1&$-4$&$-16$&$2-4R_v -16R_s$ \\
$X:= \sqrt{Y_1 Y_2^2 Y_3^3 Y_4^4 Y_5^2 Y_6^2 }$&1&1&1&$-4$&$-12$&$2 -4R_v -12R_s$  \\
$V:= (Y_1 Y_2^2 Y_3^3 Y_4^4 Y_5^2 Y_6^3)^{\frac{1}{3}}$&1&1&1&$-4$&$-8$&$2-4R_v-8R_s$ \\ \hline
  \end{tabular}}
  \end{center}\label{Spin12220}
\end{table}

%%%%%%%%%%%%%%%%%%%%%%%%%%%%%%%%%%%%%%%%%%%%%%%%%%%%%%%%%%
\subsection{$(N_v,N_s,N_{s'})=(2,1,1)$}
%%%%%%%%%%%%%%%%%%%%%%%%%%%%%%%%%%%%%%%%%%%%%%%%%%%%%%%%%
The third example is the 3d $\mathcal{N}=2$ $Spin(12)$ theory with two vectors, one spinor and one conjugate spinor. The Coulomb branch $Y$ is not allowed since the low-energy $SO(10)$ theory with two vectors generates a runaway potential and its vacuum is unstable. The Coulomb branch $V$ cannot be turned on since the stability of this branch at least requires two third-order anti-symmetric tensors $\mathbf{20}_0 \in \mathbf{32}$. 
The Coulomb branch $Z$ is stable since the $SU(2)$ dynamics is made stable by the massless components of the two spinors and since the $SO(8)$ dynamics is also stable and supersymmetric by two vectors and two spinors.
The Coulomb branch $X$ is also available since the low-energy $SO(4) \times SU(4)$ dynamics can be stable due to $(\mathbf{2},\mathbf{6})_{0}$ and $(\mathbf{2}^*,\mathbf{6})_{0}$. Table \ref{Spin12211} shows the moduli fields and their quantum numbers. We will not explicitly write down the superpotential, but one can do it from Table \ref{Spin12211}.

\begin{table}[H]\caption{3d $\mathcal{N}=2$ $Spin(12)$ theory with $(N_v,N_s,N_{s'})=(2,1,1)$} 
\begin{center}
\scalebox{0.9}{
  \begin{tabular}{|c||c||c|c|c|c|c| } \hline
  &$Spin(12)$&$SU(2)$&$U(1)_v$&$U(1)_s$&$U(1)_{s'}$&$U(1)_R$ \\ \hline
Q& $\mathbf{12}$&${\tiny \yng(1)}$&1&0&0& $R_v$ \\
$S$ & $\mathbf{32}$&1&0&1&0& $R_s$ \\ 
$S'$ & $\mathbf{32'}$&1&0&0&1& $R_{s'}$ \\  \hline
$M_{QQ}:=QQ$&1&${\tiny \yng(2)}$&2&0&0&$2R_v$ \\
$P_2:=SQ^2S$&1&1&2&2&0&$2R_v + 2R_s$  \\
$P'_2:=S'Q^2S'$&1&1&2&0&2&$2R_v + 2R_{s'}$  \\
$M_{1,SS'}:=SQS'$&1&${\tiny \yng(1)}$&1&2&2&$R_v +R_s +R_{s'}$  \\
$B:=S^4$&1&1&0&4&0&$R_s$ \\
$B':={S'}^4$&1&1&0&0&4&$4R_{s'}$ \\
$F_0:=S^2S'^2$&1&1&0&2&2&$2R_s +2R_{s'}$ \\
$F_2:=S^2S'^2Q^2$&1&${\tiny \yng(2)}$&2&2&2&$2R_v+2R_s +2R_{s'}$ \\
$F'_2:=S^2S'^2Q^2$&1&1&2&2&2&$2R_v+2R_s +2R_{s'}$ \\
$C:=S^3S' Q$&1&${\tiny \yng(1)}$&1&3&1&$R_v+3R_s+R_{s'}$ \\
$C:=S{S'}^3 Q$&1&${\tiny \yng(1)}$&1&1&3&$R_v+3R_s+R_{s'}$ \\
$T:=S^3S'^3Q$&1&${\tiny \yng(1)}$&1&3&3&$R_v+3R_s+3R_{s'}$ \\
$D:=S^4S'^2Q^2$&1&1&2&4&2&$2R_v+4R_s+2R_{s'}$ \\
$D':=S2S'^4Q^2$&1&1&2&2&4&$2R_v+2R_s+4R_{s'}$ \\
$U_0:=S^4S'^4$&1&1&0&4&2&$4R_s+4R_{s'}$ \\
$U_2:=S^4S'^4Q^2$&1&1&2&4&4&$2R_v+4R_s+4R_{s'}$ \\ \hline
%$P_6:=SQ^6S$&1&1&6&2&$6R_v + 2R_s$  \\
%$B:=S^4$&1&1&0&4&$4R_s$ \\
%$F:=S^4Q^6$&1&1&6&4&$6R_v +4R_s$ \\ \hline 
$Z:=Y_1 Y_2^2 Y_3^2 Y_4^2 Y_5 Y_6 $&1&1&$-4$&$-8$&$-8$&$2-4R_v -8R_s-8R_{s'}$ \\ 
$X:= \sqrt{Y_1 Y_2^2 Y_3^3 Y_4^4 Y_5^2 Y_6^2 }$&1&1&$-4$&$-6$&$-6$&$2 -4R_v -6R_s-6R_{s'}$  \\ \hline
  \end{tabular}}
  \end{center}\label{Spin12211}
\end{table}

%%%%%%%%%%%%%%%%%%%%%%%%%%%%%%%%%%%%%%%%%%%%%%%%%%%%%%%%%%
%%%%%%%%%%%%%%%%%%%%%%%%%%%%%%%%%%%%%%%%%%%%%%%%%%%%%%%%%%
%%%%%%%%%%%%%%%%%%%%%%%%%%%%%%%%%%%%%%%%%%%%%%%%%%%%%%%%%%
\section{$Spin(13)$ theories}
%%%%%%%%%%%%%%%%%%%%%%%%%%%%%%%%%%%%%%%%%%%%%%%%%%%%%%%%%%
%%%%%%%%%%%%%%%%%%%%%%%%%%%%%%%%%%%%%%%%%%%%%%%%%%%%%%%%%%
%%%%%%%%%%%%%%%%%%%%%%%%%%%%%%%%%%%%%%%%%%%%%%%%%%%%%%%%%%
Let us study the Coulomb branch of the 3d $\mathcal{N}=2$ $Spin(13)$ gauge theory with $N_v$ vectors and $N_s$ spinors whose dimension is $\mathbf{64}$. There are a lot of classical Coulomb branches but most of them are quantum-mechanically excluded from the quantum moduli space since almost all the components of the matter fields are massive along those directions and we will obtain the pure SYM or SQCD as a low-energy description, which will not have enough charged matters to make the SUSY vacuum stable. Therefore, we are left with a few Coulomb branch directions. 

The first candidate denoted as $Y$ corresponds to the breaking
\begin{align}
so(13) & \rightarrow so(11) \times u(1)  \\
\mathbf{13} & \rightarrow  \mathbf{11}_0+\mathbf{1}_{2} +\mathbf{1}_{-2}  \\
\mathbf{64} & \rightarrow  \mathbf{32}_1 +  \mathbf{32}_{-1}.
\end{align}
All the components of the spinor representations are massive and integrated out along this branch while the vector matter reduces to a massless vector $\mathbf{11}_0$. Therefore, the moduli space of the $Spin(13)$ theory only with spinors cannot have this operator. In order to make the low-energy $SO(11)$ vacuum stable, there must be more than eight vector matters.

The second Coulomb branch $Z$ breaks the gauge group as 
\begin{align}
so(13) & \rightarrow so(9) \times su(2) \times u(1) \\
\mathbf{13} & \rightarrow (\mathbf{9}, \mathbf{1})_{0}+(\mathbf{1}, \mathbf{2})_{\pm 1} \\
\mathbf{64} & \rightarrow (\mathbf{16}, \mathbf{2})_{0} +(\mathbf{16}, \mathbf{1})_{\pm 1}.
\end{align}
The vacuum of the low-energy $SU(2)$ theory is made stable by the massless component $(\mathbf{16}, \mathbf{2})_{0} \in \mathbf{64}$ while the $SO(9)$ part can have a stable SUSY vacuum via $(\mathbf{9}, \mathbf{1})_{0} \in \mathbf{13}$ or $(\mathbf{16}, \mathbf{2})_{0} \in \mathbf{64}$. Therefore, the $Spin(13)$ theory with spinor matters includes this branch.

The third candidate denoted as $X$ corresponds to the breaking
\begin{align}
so(13) & \rightarrow so(5) \times su(4) \times u(1) \\
\mathbf{13} & \rightarrow  (\mathbf{5}, \mathbf{1})_{0}+(\mathbf{1}, \mathbf{4})_{1}+(\mathbf{1}, \overline{\mathbf{4}})_{-1} \\
\mathbf{64} & \rightarrow (\mathbf{4}, \mathbf{6})_{0} +(\mathbf{4}, \mathbf{1})_{2} +(\mathbf{4}, \mathbf{1})_{-2} +(\mathbf{4}, \mathbf{4})_{-1} +(\mathbf{4}, \overline{\mathbf{4}})_{1}.
\end{align}
The vector field cannot make the low-energy $SU(4)$ vacuum stable since there is no massless component charged under the $SU(4)$ subgroup. When the theory has at least one spinor, the component $(\mathbf{4}, \mathbf{6})_{0} \in \mathbf{64}$ makes the $SO(5) \times SU(4)$ dynamics stable and keeps it supersymmetric. Therefore, the $Spin(13)$ theory with spinor matters also includes this operator.

Finally, we mention that there could be an additional Coulomb branch operator $V$ which corresponds to the breaking
\begin{align}
so(13) & \rightarrow  su(6) \times u(1) \\
\mathbf{13} & \rightarrow \mathbf{1}_0 +\mathbf{6}_1 + \overline{\mathbf{6}}_{-1}  \\
\mathbf{64} & \rightarrow  \mathbf{20}_0+ \mathbf{6}_{-2}  +\overline{\mathbf{6}}_{2}  +\mathbf{15}_{-1}+\overline{\mathbf{15}}_{1} +\mathbf{1}_3 +\mathbf{1}_{-3}. 
\end{align}
Along this direction, the massless components $\mathbf{20}_0 \in \mathbf{64} $ can make the $SU(6)$ vacuum stable. However, this is only possible when there are two spinors in the theory. In what follows. we will only consider the $Spin(13)$ theory with a single spinor and this operator is not necessary.

%%%%%%%%%%%%%%%%%%%%%%%%%%%%%%%%%%%%%%%%%%%%%%%%%%%%%%%%%%
\subsection{$(N_v,N_s)=(3,1)$}
%%%%%%%%%%%%%%%%%%%%%%%%%%%%%%%%%%%%%%%%%%%%%%%%%%%%%%%%%%
The 3d $\mathcal{N}=2$ $Spin(13)$ theory with three vectors and one spinor exhibits s-confinement. The Higgs branch of the moduli space of vacua is described by eleven composite operators $M_{QQ}, P_2, P_3, R_0, R_1, R_2, R_3, T_2, T_3, U_0$ and $U_3$, which are defined in Table \ref{Spin1331}. The Coulomb branch is two-dimensional and this is described by $Z$ and $X$ which are defined above. We will not show the confining superpotential since the explicit form is cumbersome. Table \ref{Spin1331} summarizes the quantum numbers of the moduli operators.

\begin{table}[H]\caption{3d $\mathcal{N}=2$ $Spin(13)$ theory with $(N_v,N_s)=(3,1)$} 
\begin{center}
\scalebox{1}{
  \begin{tabular}{|c||c||c|c|c|c| } \hline
  &$Spin(13)$&$SU(3)$&$U(1)_v$&$U(1)_s$&$U(1)_R$ \\ \hline
Q& $\mathbf{13}$&${\tiny \yng(1)}$&1&0& $R_v$ \\
$S$ & $\mathbf{64}$&1&0&1& $R_s$ \\ \hline 
$M_{QQ}:=QQ$&1&${\tiny \yng(2)}$&2&0&$2R_v$ \\
$P_2:=SQ^2S$&1&${\tiny \overline{ \yng(1)} }$&2&2&$2R_v + 2R_s$  \\
$P_3:=SQ^3S$&1&1&3&2&$3R_v + 2R_s$  \\
$R_{0}:=S^4$ &1&1&0&4&$4R_s$  \\
$R_{1}:= S^4 Q$&1&${\tiny \yng(1)}$&1&4&$R_v+4R_s$  \\
$R_{2}:=S^4 Q^2$&1&${\tiny \yng(2)}$&2&4&$2R_v+4R_s$ \\
$R_{3}:=S^4 Q^3$&1&1&3&4&$3R_v+4R_s$ \\
$T_{2}:=S^6Q^2$&1&${\tiny \overline{ \yng(1)} }$&2&6&$2R_v+6R_s$ \\
$T_{3}:=S^6Q^3$&1&1&3&6&$3R_v+6R_s$ \\
$U_{0}:=S^8$&1&1&0&8&$8R_s$ \\
$U_{3}:=S^8Q^3$&1&1&3&8&$3R_v +8R_s$ \\ \hline 
$Z:=Y_1 Y_2^2 Y_3^2 Y_4^2 Y_5^2 Y_6 $&1&1&$-6$&$-16$&$2-6R_v -16R_s$ \\
$X:= \sqrt{Y_1 Y_2^2 Y_3^3 Y_4^4 Y_5^4 Y_6^2 }$&1&1&$-6$&$-12$&$2 -6R_v -12R_s$  \\ \hline
  \end{tabular}}
  \end{center}\label{Spin1331}
\end{table}

%%%%%%%%%%%%%%%%%%%%%%%%%%%%%%%%%%%%%%%%%%%%%%%%%%%%%%%%%%
%%%%%%%%%%%%%%%%%%%%%%%%%%%%%%%%%%%%%%%%%%%%%%%%%%%%%%%%%%
%%%%%%%%%%%%%%%%%%%%%%%%%%%%%%%%%%%%%%%%%%%%%%%%%%%%%%%%%%
\section{$Spin(14)$ theories}
%%%%%%%%%%%%%%%%%%%%%%%%%%%%%%%%%%%%%%%%%%%%%%%%%%%%%%%%%%
%%%%%%%%%%%%%%%%%%%%%%%%%%%%%%%%%%%%%%%%%%%%%%%%%%%%%%%%%%
%%%%%%%%%%%%%%%%%%%%%%%%%%%%%%%%%%%%%%%%%%%%%%%%%%%%%%%%%%
The final example is the 3d $\mathcal{N}=2$ $Spin(14)$ gauge theory with vector ($\mathbf{14}$) matters and one spinor ($\mathbf{64}$). Since the dimension of the spinor representation is huge, the theories with more than one spinor ($\mathbf{64}$ or $\overline{\mathbf{64}}$) will exhibit a conformal window or a non-abelian Coulomb phase. Since we are now interested in the s-confinement phases of the $Spin(N)$ gauge theories, we focus on the $Spin(14)$ theory with one spinor and some vectors. 
 
There are two Coulomb branches which we have to take into account. The vev of the first coordinate $Z$ corresponds to the breaking 
\begin{align}
so(14) &\rightarrow so(10) \times su(2) \times u(1) \\
\mathbf{14}  &\rightarrow (\mathbf{10},\mathbf{1}  )_0 + (\mathbf{1},\mathbf{2} )_{\pm 1} \\
\mathbf{64}  &  \rightarrow (\mathbf{16}, \mathbf{1} )_{\pm 1} +(\overline{\mathbf{16}},\mathbf{2}   )_0.
\end{align}
The Chern-Simons term for $U(1)$ is not introduced as it should be. This is a necessary condition that the Coulomb branch $Z$ can be a flat direction. The resulting low-energy theory contains the 3d $\mathcal{N}=2$ $SO(10) \times SU(2)$ SQCD with massless chiral superfields in fundamental and spinor representations of $SO(10)$ and $SU(2)$. In order that the coordinate $Z$ can be a stable vacuum, there must be enough matters charged under the $SO(10) \times SU(2)$. For example, the theories without a spinor matter cannot have this flat direction since there is no massless field charged under the $SU(2)$ and then the $SU(2)$ vacuum is unstable due to the monopole potential. 

The second Coulomb branch $X$ corresponds to the breaking
\begin{align}
so(14) &\rightarrow so(6) \times su(4) \times u(1) \\
\mathbf{14}  &\rightarrow (\mathbf{6},\mathbf{1}  )_0+(\mathbf{1},\mathbf{4}  )_1 +(\mathbf{1},\overline{\mathbf{4}}  )_{-1} \\
\mathbf{64}  &  \rightarrow (\mathbf{4},\mathbf{4}  )_{-1} +(\mathbf{4},  \overline{\mathbf{4} } )_1+(\overline{\mathbf{4}},\mathbf{6}  )_0+(\overline{\mathbf{4}},\mathbf{1}  )_2 +( \overline{\mathbf{4}},\mathbf{1}  )_{-2}.
\end{align}
This vacuum can be stable by massless components $(\mathbf{6},\mathbf{1}  )_0$ or $(\overline{\mathbf{4}},\mathbf{6}  )_0$. The theory only with vector matters cannot include this operator since the low-energy $SU(4)$ dynamics is unstable. For the theory with a spinor matter, each gauge dynamics can be stable due to $(\overline{\mathbf{4}},\mathbf{6}  )_0$.

Notice that when we introduce more general matter contents ($N_v$ vectors, $N_s$ spinors and $N_{s'}$ complex conjugate spinors), there may be additional Coulomb branches. For instance, the classical Coulomb branch will include the following direction 
\begin{align}
so(14) &\rightarrow  su(6) \times so(2)  \times u(1) \\
\mathbf{14}  &\rightarrow  \mathbf{6}_{0,1} + \overline{\mathbf{6}}_{0,-1} +\mathbf{1}_{2,0} +\mathbf{1}_{-2,0} \\
\mathbf{64}  &\rightarrow \mathbf{6}_{1,-2}+\mathbf{20}_{1,0} + \overline{\mathbf{6}}_{1,2} +\mathbf{15}_{-1,-1} + \overline{\mathbf{15}}_{-1,1} +\mathbf{1}_{-1,3}+\mathbf{1}_{-1,-3} \\
\overline{\mathbf{64}}  &\rightarrow \mathbf{6}_{-1,-2}+\mathbf{20}_{-1,0} + \overline{\mathbf{6}}_{-1,2} +\mathbf{15}_{1,-1} + \overline{\mathbf{15}}_{1,1} +\mathbf{1}_{-1,3}+\mathbf{1}_{1,-3},
\end{align}
where the Coulomb branch operator corresponds to the second $U(1)$ factor. We can use two massless components $\mathbf{20}_{\pm 1,0} $ in order to have a stable vacuum of the low-energy $SU(6)$ gauge theory. When the theory has $N_v \ge 10$ vector matters, there is another Coulomb branch $Y$ which corresponds to the breaking
\begin{align}
so(14) &\rightarrow  so(12) \times u(1) \\
\mathbf{14}  &\rightarrow \mathbf{12}_0 +\mathbf{1}_2 +\mathbf{1}_{-2} \\
\mathbf{64}  &\rightarrow \mathbf{32}_{1}+\mathbf{32'}_{-1}.
\end{align}
All the components of the spinor are massive and only the vector matters can make the low-energy $SO(12)$ theory stable.

%%%%%%%%%%%%%%%%%%%%%%%%%%%%%%%%%%%%%%%%%%%%%%%%%%%%%%%%%%
\subsection{$(N_v,N_s,N_{s'})=(4,1,0)$}
%%%%%%%%%%%%%%%%%%%%%%%%%%%%%%%%%%%%%%%%%%%%%%%%%%%%%%%%%
The s-confinement phase appears only in the 3d $\mathcal{N}=2$ $Spin(14)$ gauge theory with four vectors and one spinor (or four vectors and one conjugate spinor). The Higgs branch is described by seven composites: $M_{QQ}, P_3, B_{4,2}, B_{4,4}, B_{6,3}, B_{8,0}$ and $B_{8,4}$ defined in Table \ref{Spin1441}. As explained above, there are two Coulomb branch coordinates $Z$ and $X$. The superpotential takes
\begin{align}
W &= Z \left[B_{8,0}^2 \det  M_{QQ}+ \det B_{4,2} +B_{8,0} (M_{QQ}^2 B_{4,2}^2 +M_{QQ}P_3 B_{6,3} +B_{4,2} P_3^2 )  \right.  \nonumber \\ &\qquad   \left. +B_{4,2} B_{6,3}^2 +B_{4,4}^2 B_{8,0} +B_{8,4}^2 \right] \nonumber \\ 
& \quad +X \left[ (M_{QQ}^3 B_{4,2}+M_{QQ}P_3^2) B_{8,0} +M_{QQ}( B_{4,2}^3 +B_{6,3}^2)+ P_3 B_{6,3} B_{4,2} +B_{4,4} B_{8,4}  \right].
\end{align}
\begin{table}[H]\caption{3d $\mathcal{N}=2$ $Spin(14)$ theory with $(N_v,N_s,N_{s'})=(4,1,0)$} 
\begin{center}
\scalebox{1}{
  \begin{tabular}{|c||c||c|c|c|c| } \hline
  &$Spin(14)$&$SU(4)$&$U(1)_v$&$U(1)_s$&$U(1)_R$ \\ \hline
Q& $\mathbf{14}$&${\tiny \yng(1)}$&1&0& $R_v$ \\
$S$ & $\mathbf{64}$&1&0&1& $R_s$ \\ \hline 
$M_{QQ}:=QQ$&1&${\tiny \yng(2)}$&2&0&$2R_v$ \\
$P_3:=SQ^3S$&1&${\tiny \overline{ \yng(1)} }$&3&2&$3R_v + 2R_s$  \\
$B_{4,2}:=S^4Q^2$ &1&${\tiny \yng(2)}$&2&4&$2R_v +4R_s$  \\
$B_{4,4}:= S^4 Q^4$&1&1&4&4&$4R_v+4R_s$  \\
$B_{6,3}:=S^6 Q^3$&1&${\tiny \overline{\yng(1)}}$&3&6&$3R_v+6R_s$ \\
$B_{8,0}:=S^8$&1&1&0&8&$8R_s$ \\
$B_{8,4}:=S^8Q^4$&1&1&4&8&$4R_v +8R_s$ \\ \hline 
$Z:=Y_1 Y_2^2 Y_3^2 Y_4^2 Y_5^2 Y_6 Y_7 $&1&1&$-8$&$-16$&$2-8R_v -16R_s$ \\
$X:= \sqrt{Y_1 Y_2^2 Y_3^3 Y_4^4 Y_5^4 Y_6^2 Y_7^2}$&1&1&$-8$&$-12$&$2 -8R_v -12R_s$  \\ \hline
  \end{tabular}}
  \end{center}\label{Spin1441}
\end{table}

%%%%%%%%%%%%%%%%%%%%%%%%%%%%%%%%%%%%%%%%%%%%%%%%%%%%%%%%%%
%%%%%%%%%%%%%%%%%%%%%%%%%%%%%%%%%%%%%%%%%%%%%%%%%%%%%%%%%%
%%%%%%%%%%%%%%%%%%%%%%%%%%%%%%%%%%%%%%%%%%%%%%%%%%%%%%%%%%
\section{Summary}
%%%%%%%%%%%%%%%%%%%%%%%%%%%%%%%%%%%%%%%%%%%%%%%%%%%%%%%%%%
%%%%%%%%%%%%%%%%%%%%%%%%%%%%%%%%%%%%%%%%%%%%%%%%%%%%%%%%%%
%%%%%%%%%%%%%%%%%%%%%%%%%%%%%%%%%%%%%%%%%%%%%%%%%%%%%%%%%%

%[What we have done]
In this paper, we investigated the various s-confinement phases in the 3d $\mathcal{N}=2$ $Spin(N)$ gauge theories with vector matters and spinor matters. We found that the 3d s-confinement is connected with the (quantum-deformed) moduli space of the corresponding 4d $\mathcal{N}=1$ $Spin(N)$ gauge theories via the twisted-monopole superpotential \cite{Grinstein:1997zv, Grinstein:1998bu}. Naively, one might consider that almost all the classical Coulomb branches are quantum-mechanically lifted since the matter fields are massive and the non-perturbative superpotential lifts those flat directions. However, we pointed out that the $Spin(N)$ theory with vectors and spinors can have the additional Coulomb branches. Along these new branches, some components of the spinor fields can survive as massless fields and they can make these flat directions stable and supersymmetric. 

We gave a systematic study of the Coulomb branch and the s-confinement phases for the 3d $\mathcal{N}=2$ $Spin(N)$ gauge theories. Although the analysis of the Coulomb branch was systematic, the resulting Coulomb branch structure was changing, depending on the rank of the gauge group. For example, the $Spin(10)$ theory with three spinors and one conjugate spinor was very special and we needed to introduce the ``dressed'' Coulomb branch operator. This was because there are two unbroken $U(1)$ subgroups along the Coulomb branch and the mixed Chern-Simons term is introduced. As another example, the $Spin(12)$ theory with two vectors and two spinors exhibited the three-dimensional Coulomb branch while, in most other cases, the Coulomb branch was one- or two-dimensional.

%[remaining problems]
Since we are interested in the s-confinement phases, the number of the spinor matters is highly restricted especially in the case of large $Spin(N)$ gauge groups where the dimensions of the spinors are huge. When there are more spinor matters, we could define the additional Coulomb branches which survive quantum corrections. Remember the two examples, $Spin(13)$ and $Spin(14)$, where we claimed that the additional Coulomb branch will be necessary when there are more than one spinor. It is important to check the validity of this analysis, for instance, by computing the superconformal indices  \cite{Bhattacharya:2008bja, Kim:2009wb}. This is a hard and challenging problem since the rank of the gauge group is large and the calculation would be quite heavy.

%[Future directions]
In this paper, we focused on the s-confinement phases of the 3d $Spin(N)$ gauge theory and proposed various confining phases. It is important to test our proposal, for instance, by computing the superconformal indices \cite{Bhattacharya:2008bja, Kim:2009wb}. The dual descriptions are given by the non-gauge theories presented here. It is also very important to study different phases by more introducing vector and spinor matters. For example, the Seiberg dualities of the 4d $Spin(N)$ theories were studied in \cite{Pouliot:1995zc, Pouliot:1995sk, Cho:1997kr, Pouliot:1996zh, Kawano:1996bd, Berkooz:1997bb, Kawano:2005nc}. One can, in principle, derive the corresponding 3d dualities from the 4d ones by following the argument in \cite{Aharony:2013dha, Aharony:2013kma}. We will soon come back to this problem elsewhere.

%%%%%%%%%%%%%%%%%%%%%%%%%%%%%%%%%%%%%%%%%%%%%%%%%%%%%%%%%%
%%%%%%%%%%%%%%%%%%%%%%%%%%%%%%%%%%%%%%%%%%%%%%%%%%%%%%%%%%
\section*{Acknowledgments}
%%%%%%%%%%%%%%%%%%%%%%%%%%%%%%%%%%%%%%%%%%%%%%%%%%%%%%%%%%
%%%%%%%%%%%%%%%%%%%%%%%%%%%%%%%%%%%%%%%%%%%%%%%%%%%%%%%%%%
I would like to thank Prof. Jan de Boer for helpful discussion.
This work is supported by the Swiss National Science Foundation (SNF) under grant number PP00P2\_157571/1 and by ``The Mathematics of Physics'' (SwissMAP) under grant number NCCR 51NF40-141869.

\bibliographystyle{ieeetr}
\bibliography{ConfinementSpinRef}

\end{document}